\documentclass[twocolumn,times]{aastex62}

\usepackage{natbib}
\usepackage{amsmath}
\usepackage{textcomp}
\usepackage{graphicx}
\usepackage{ifpdf}
\usepackage[T1]{fontenc}
\usepackage{atbegshi}
\usepackage{amssymb}
\usepackage{dcolumn}

\newcommand{\msun}{$M_{\odot}$}

\newcommand{\zsun}{$Z_{\odot}$}

\def\tol[#1][#2][#3]{${#1}^{+{#2}}_{-{#3}}$}

\begin{document}

\title{Mapping the Escape Fraction of Ionizing Photons Using Resolved Stars: A Much Higher Escape Fraction for NGC 4214}

\correspondingauthor{Yumi Choi}
\email{ychoi@stsci.edu}

\author{Yumi Choi}
\affiliation{Space Telescope Science Institute, 3700 San Martin Drive, Baltimore, MD 21218, USA}

\author{Julianne J. Dalcanton}
\affiliation{Department of Astronomy, University of Washington, Box 351580, Seattle, WA 98195, USA}

\author{Benjamin F. Williams}
\affiliation{Department of Astronomy, University of Washington, Box 351580, Seattle, WA 98195, USA}

\author{Evan D. Skillman}
\affiliation{Minnesota Institute for Astrophysics, University of Minnesota, 116 Church Street SE, Minneapolis, MN 55455, USA}

\author{Morgan Fouesneau}
\affiliation{Max Planck Institut f\"ur Astronomie, K\"onigstuhl 17, D-69117 Heidelberg, Germany}

\author{Karl D. Gordon}
\affiliation{Space Telescope Science Institute, 3700 San Martin Drive, Baltimore, MD 21218, USA}

\author{Karin M. Sandstrom}
\affiliation{Center for Astrophysics and Space Sciences, University of California, San Diego, 9500 Gilman Drive, La Jolla, CA 92093, USA}

\author{Daniel R. Weisz}
\affiliation{Astronomy Department, University of California, Berkeley, CA 94720, USA}

\author{Karoline M. Gilbert}
\affiliation{Space Telescope Science Institute, 3700 San Martin Drive, Baltimore, MD 21218, USA}

\shortauthors{Choi et al.}

\begin{abstract}
We demonstrate a new method for measuring the escape fraction of ionizing photons using Hubble Space Telescope imaging of resolved stars in NGC 4214, a local analog of high-redshift starburst galaxies that are thought to be responsible for cosmic reionization. Specifically, we forward model the UV through near-IR spectral energy distributions of $\sim$83,000 resolved stars to infer their individual ionizing flux outputs. We constrain the local escape fraction by comparing the number of ionizing photons produced by stars to the number that are either absorbed by dust or consumed by ionizing the surrounding neutral hydrogen in individual star-forming regions. We find substantial spatial variation in the escape fraction (0--40\%). Integrating over the entire galaxy yields a global escape fraction of \tol[25][16][15]\%. This value is much higher than previous escape fractions of zero reported for this galaxy. We discuss sources of this apparent tension, and demonstrate that the viewing angle and the 3D ISM geometric effects are the cause. If we assume the NGC 4214 has no internal dust, like many high-redshift galaxies, we find an escape fraction of 59\% (an upper-limit for NGC 4214). This is the first non-zero escape fraction measurement for UV-faint (M$_{\rm FUV}$ = --15.9) galaxies at any redshift, and supports the idea that starburst UV-faint dwarf galaxies can provide a sufficient amount of ionizing photons to the intergalactic medium.
\end{abstract}


\section{Introduction} \label{sec:intro}
Cosmic reionization is one of the major events in the evolution of our Universe. Observations of the Cosmic Microwave Background (CMB), Ly$\alpha$ forest absorption in quasar spectra, and the rapid decrease in number density of Ly$\alpha$ emitters indicate that hydrogen in the neutral intergalactic medium (IGM) became completely ionized by a redshift ($z$) of 6 \citep{gunn65, fan06, dunkley09, ouchi10, stark10, ono12, schenker12, becker15, mcgreer15, planck16}. Until very recently, there was no consensus on the main contributors to cosmic reionization: low luminosity star-forming galaxies \citep[e.g.,][]{barkana01, wise09, kuhlen12, boylankolchin14, wise14, anderson17} versus active galactic nuclei \citep[e.g.,][]{kollmeier14, giallongo15, madau15}. Pushing the rest-frame ultraviolet (UV) observational frontier up to $z \simeq$ 9--12 has shed light on our understanding of the role of faint star-forming galaxies in cosmic reionization, leading to a growing belief that ionizing (LyC) photons escaping from numerous faint star-forming galaxies might be sufficient to reionize the early universe \citep[e.g.,][]{bouwens12, oesch13, boylankolchin14, bouwens15, robertson15, finkelstein15, mcleod16, livermore17, ishigaki18}. 

In this scenario, faint star-forming galaxies release LyC photons through low-density ``holes'' in their clumpy interstellar medium (ISM), which can be created by strong stellar feedback from the most massive stars \citep[e.g.,][]{heckman11, jaskot13, zastrow13, alexandroff15}. The escaped LyC photons then ionize the surrounding neutral IGM. Confirming whether faint star-forming galaxies were indeed responsible for cosmic reionization requires constraining the fraction of LyC photons that escape (the ``escape fraction'', $f_{\rm esc}$) into the IGM. Unfortunately, it is impossible to directly detect escaping LyC photons from galaxies at the epoch of reionization due to absorption by the high neutral fraction of  the IGM. Therefore, looking at lower-$z$ analogs is currently the only way to constrain the physics that controls the escape fraction.

\begin{figure*}
\centering
\vspace{-2,5cm}
\includegraphics[width=18cm]{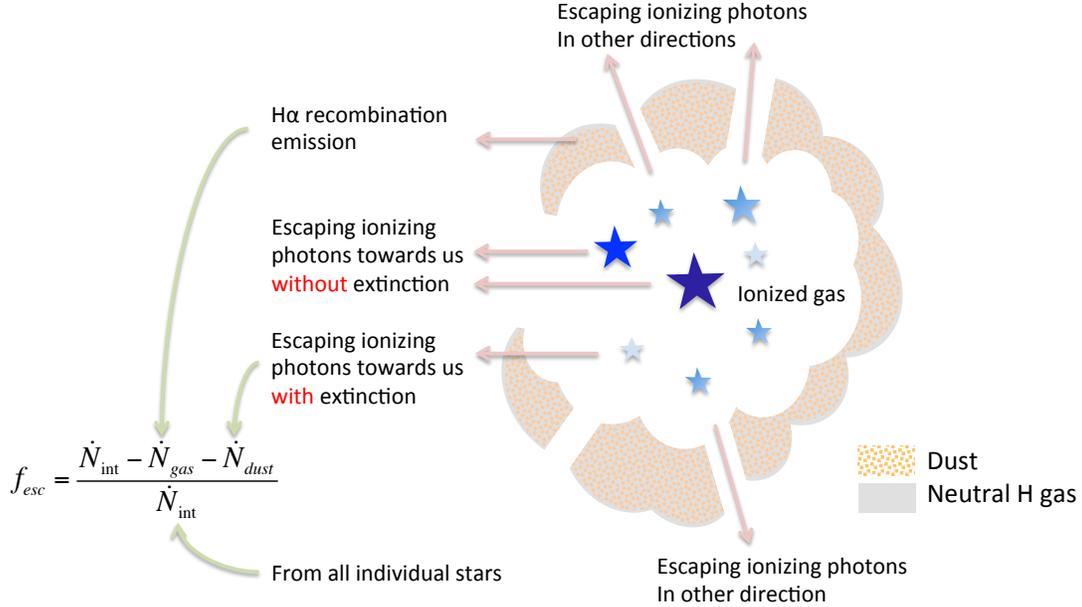}
\vspace{-3cm}
\caption{Schematic of our method for deriving the escape fraction of ionizing photons by measuring the (1) intrinsic rate of production of ionizing photons ($\dot{N}_{\rm int}$) from SED fitting, (2) rate of consuming ionizing photons by neutral hydrogen ($\dot{N}_{\rm gas}$) from the H$\alpha$ recombination line, and (3) rate of absorption of ionizing photons by dust ($\dot{N}_{\rm dust}$) from the same SED fitting. The effect of the relative geometry between stars and dust can be taken into account by introducing the covering factor, the fraction of HII surface covered by the dust. We note that the gas/dust distribution presented here is for illustration purpose only. 
\label{cartoonEscape}}
\end{figure*}

\subsection{Current Constraints on $f_{\rm esc}$}
Although a general consensus has been reached that low luminosity star-forming galaxies dominate cosmic reionization, no $f_{\rm esc}$ measurements have been made for observed star-forming galaxies fainter than $M_{\rm UV} \simeq$ --19 yet. Thus, observational constraints on $f_{\rm esc}$ for UV-faint galaxies are urgent, especially at --19 $ < M_{\rm UV} <$ --10. However, the actual escape fractions have been difficult to constrain in both simulations and observations. 

To be consistent with reionizing the Universe by redshift $z \sim$ 6, $f_{\rm esc}$ must be at least 10--30\%, depending on the choice of the ionizing photon production efficiency and the faint-end properties (i.e., slope and the limiting magnitude) of the observed rest-frame UV luminosity functions \citep[e.g.,][]{ouchi09, finkelstein12, robertson13, bouwens15, khaire16, ishigaki18}. Theory offers few constrains on $f_{\rm esc}$, given that state-of-the art models predict a wide range of $f_{\rm esc}$ (0--100\%) and different trends in redshift evolution \citep[e.g.,][]{wood00, fujita03, gnedin08, wise09, razoumov10, yajima11, kim13, kimm14, ma15, paardekooper15, xu16, sharma17}. 

The observational situation is equally ambiguous. Despite many attempts to detect escaping ionizing photons from galaxies, $f_{\rm esc}$ has been poorly constrained by observations at any redshift (0.04 $\lesssim z \lesssim$ 5) and only a small fraction of galaxies out of many candidates show any evidence for leakage with most at a level of a few per cent (up to $\sim$10\%) \citep[e.g.,][]{leitherer95, heckman01, bergvall06, inoue06, shapley06, siana07, grimes09, iwata09, vanzella12, leitet13, nestor13, borthakur14, izotov16a, izotov16b, leitherer16,  vasei16, rutkowski17}. $f_{\rm esc}$ at a level of $>$ 20\% has been measured only very recently for a small number of galaxies based on direct detection of leaking LyC flux \citep[e.g.,][]{vanzella12, shapley16,vanzella16,bian17,naidu17,izotov18}. 

Even this handful of $f_{\rm esc}$ measurements have proven difficult to interpret. These low $f_{\rm esc}$ measurements and/or a low direct detection rate of LyC leakers may be biased due, in part, to sample selection favoring relatively massive galaxies or strong H$\alpha$ emission-line galaxies (i.e., high H$\,\textsc{i}$ column density, thus insignificant leakage of ionizing photons). Instead, it has been suggested that using the line ratio of [O$\,\textsc{iii}$]$\lambda$5007/[O$\,\textsc{ii}$]$\lambda$3727 (the O32 ratio, a proxy for the ionization parameter) is more efficient to select actual LyC leakers \citep[e.g.,][]{jaskot13,izotov16b, izotov18b}, but the correlation between the O32 ratio and $f_{\rm esc}$ is still a matter of debate \citep[see][]{naidu18, bassett19}. Strong O32 galaxies tend to emit strong Ly$\alpha$ emission \cite[e.g.][]{hayes14,nakajima14,ostlin14}. These emission line properties of the LyC leakers and their correlation with $f_{\rm esc}$ have advanced our understanding on the ionizing photon escape mechanism \citep[e.g.,][]{hayes14,dijkstra16,chisholm17,riverathorsen17,izotov18b,kakiichi19, kimm19,bian20}. Although the correlations between $f_{\rm esc}$ and nebular lines or galactic properties are not clear yet, ongoing efforts to compile a large sample of true LyC leakers will make it possible to reveal the LyC escaping mechanism and establish the role of SF galaxies in cosmic reionization \citep[e.g.][]{oesch18,fletcher19}. 

Another explanation for low $f_{\rm esc}$ measurements is inherent difficulties with the direct measurement of escaping ionizing photons. These include contamination from low-$z$ intervening galaxies, uncertain IGM transmission, large uncertainty in UV background subtraction, and, last but not least, narrow opening angles of optically thin holes that are misaligned with our line of sight (i.e., lowering the chance to detect leaked ionizing photons). 

Of the possible limitations, the most fundamental issue may be lack of sufficient spatial resolution to capture the local variation of $f_{\rm esc}$ within a galaxy. Constraining $f_{\rm esc}$ requires measuring both (1) the intrinsic ionizing photon production rate, and (2) either the photon absorption rate by the ISM or the amount of leaked ionizing photons. However, because ionizing photons produced by clustered O/B stars must propagate through the complex, dusty ISM \citep[e.g.,][]{witt92,witt96, dove00} before eventually escaping to the IGM, all of quantities needed to measure $f_{\rm esc}$ are sensitive to the distribution of hot stars and the ISM topology, which both vary significantly with position within a galaxy. As such, the value of $f_{\rm esc}$ will also depend on local properties, and must vary spatially. 

A strong dependence of emerging spectra on the relative star/dust geometry has been well explored in the literature \citep[e.g.][]{witt92,calzetti94,witt96,gordon97,witt00,calzetti01}. Even for the same amount of dust and identical stars, a difference in the relative geometry between stars and dust can result in a significant difference in the emerging spectra, suggesting that proper dust correction for stars is critical to infer the amount of intrinsic ionizing photons \citep{mackenty00}. For example, an assumption for the simplest relative geometry between stars and dust (e.g., a uniform dust screen) for a given amount of dust predicts the maximum dust-corrected intrinsic ionizing flux, resulting in the minimum escape fraction \citep[e.g.,][]{calzetti94}. \citet{paardekooper11} explored the resolution effect in measuring $f_{\rm esc}$ by simulating the same galaxy with different mass resolutions, and showed that simulations with 10--20 times higher resolution, which better account for the ISM porosity around the stars, have more than two orders of magnitude higher $f_{\rm esc}$. They concluded that resolving the ISM on small scales is crucial for accurately determining $f_{\rm esc}$. Therefore, failure to resolve these small-scale properties can lead to large uncertainties in $f_{\rm esc}$ measurements \citep[e.g.,][]{kimm14,trebitsch17}.

In most existing studies, one directly detects the leaking ionizing radiation, if present, from an intermediate/high-$z$ starburst galaxy by measuring its flux below the rest-frame Lyman limit ($\lambda =$ 912~\AA). One then infers its intrinsic ionizing flux by modeling the galaxy's observed composite spectrum using synthetic stellar population models with internal dust \citep[e.g., Starburst99;][]{leitherer99, calzetti01}. This process necessarily involves significant spatial averaging over a galaxy that is in reality a collection of multiple clumpy star-forming regions with different physical conditions, including the ISM porosity. 

For example, an average internal dust extinction for the galaxy is usually determined based on the {\it global} Balmer decrement or on the UV slope of its composite spectrum. In the flux-weighted galaxy spectrum, both Balmer decrement and UV slope are affected by high-density regions where ionizing photons cannot easily escape. While ionizing photons are created in those high-density regions, they actually escape through low-density or dust-free holes that are created by strong stellar feedback. Thus, spatially-averaged dust corrections are likely to overestimate the dust in the regions that actually leak ionizing photons. This inevitably overpredicts the number of intrinsic ionizing photons, and thereby underestimates the global $f_{\rm esc}$. 

Furthermore, the direction of low-density holes is determined by the initial ISM geometry; stellar feedback would create low-density, dust-free holes more easily along originally optically-thin regions compared to originally optically-thick regions. Thus, only when an observer is well-aligned with the low-density holes of a galaxy can the observer directly detect any leaking LyC fluxes. Otherwise, the $f_{\rm esc}$ based on direct detection of LyC would be very low or zero, which might be the case for most previous observational attempts that ended up with very small $f_{\rm esc}$.

\subsection{The Potential of Resolved Stellar Populations}
One way to overcome these challenges is to resolve stellar populations and the ISM up to a scale of individual star-forming regions. Metal-poor starburst dwarf galaxies at $z =$ 0, in particular, provide excellent laboratories for evaluating the plausibility of early star-forming galaxies as the source of cosmic reionization. High-resolution imaging with the Hubble Space Telescope (HST) resolves individual stars in galaxies closer than $\sim$4~Mpc. Furthermore, multi-wavelength imaging with HST allows one to correct for dust star-by-star by modeling their broad stellar spectral energy distributions (SEDs), and thus infer the intrinsic LyC flux precisely. 

Recently, we introduced a stellar SED fitting technique optimized for large resolved star datasets, called the Bayesian Extinction And Stellar Tool \citep[BEAST;][]{gordon16}. Given resolved stellar photometry over a range of wavelengths, the BEAST can infer the intrinsic spectrum and intervening dust extinction for individual stars by modeling their observed stellar SEDs using the full covariance noise model accounting for observational uncertainties. These intrinsic spectra can then be applied to map intrinsic far-UV (FUV) and extreme-UV (EUV) fluxes that are impossible to observe directly. 

Figure~\ref{cartoonEscape} summarizes our strategy for measuring the ionizing photon escape fraction for a star-forming region using individual stars. In each SF region, we count the number of LyC photons produced by stars, the number of LyC photons used to ionize neutral hydrogen gas, and the number of LyC photons consumed by the dusty ISM. First, stellar SED fitting is used to infer the production rate of intrinsic ionizing photons ($\dot{N}_{\rm int}$) from individual stars, as well as the absorption rate of ionizing photons by dust along our lines of sight. Second, we compute the rate at which ionizing photons are consumed by neutral hydrogen ($\dot{N}_{\rm gas}$) using the H$\alpha$ recombination emission line, assuming Case B recombination \citep[e.g.,][]{hummer87, kennicutt98}. We then estimate the dust absorption of LyC photons ($\dot{N}_{\rm dust}$) in other directions by utilizing the covering factor of individual SF regions from \citet{hermelo13}. With these three measurements for individual SF regions, we can create a map of $f_{\rm esc}$ for a single galaxy with high spatial resolution. 

Contrary to the existing methodology that has been applied to high-$z$ galaxies, our new approach does not use a spatially-averaged dust extinction correction based on simplified assumptions about dust geometry (e.g., a uniform dust screen) to infer the intrinsic LyC production rate and the H$\alpha$ luminosity. Likewise, our method does not require challenging observations below the Lyman limit for direct detection of LyC leakage, enabling detailed study on the ionizing photon escape process in $z =$ 0 galaxies. Furthermore, our method does not require alignment between our sightlines and opening angles of porous HII regions.

In Section~\ref{sec:ngc4214}, we begin with the description of a starburst galaxy NGC 4214, an excellent analog for high-$z$ star-forming galaxies. In Section~\ref{sec:data}, we present a brief overview of its HST imaging data and photometry, and the ancillary data used in this work. In Section~\ref{n4214_setting}, we present a brief description of the generic BEAST procedure and specific BEAST settings for NGC 4214. Section~\ref{sec:results} presents the fitting results for NGC 4214. In Section~\ref{sec:fesc}, we measure the local and global escape fractions, and briefly discuss the implication for cosmic reionization and various effects in our escape fraction fraction measurement. We summarize our results and describe future work in Section~\ref{sec:summary}. 

We adopt a distance of 3.04~Mpc \citep{dalcanton09}. With this distance, an angular separation of 1$''$ corresponds to 14.7~pc.

\section{NGC 4214 as a Test Case}\label{sec:ngc4214}
Herein, we apply our new technique to resolved photometry of NGC 4214, a metal-poor starburst dwarf galaxy \citep[M$_{\rm Star} \simeq$ 1.5$\times10^{9}$~\msun;][]{karachentsev04} with $M_{\rm FUV} =$ --15.9 \citep{lee11}. Such nearby analogs of high-$z$ star-forming galaxies are excellent laboratories to study the star formation (SF) process, and the impact of stellar feedback on the ISM and the surrounding IGM at low metallicities. NGC 4214 is an especially appropriate galaxy to measure the escape fraction because of the presence of superbubbles in the galactic center, along which ionizing photons can readily escape into the IGM. Furthermore, this study will provide the first robust escape fraction measurement for dwarf galaxies fainter than $M_{\rm FUV} =$ --19.    

\begin{figure}
\includegraphics[width=9cm]{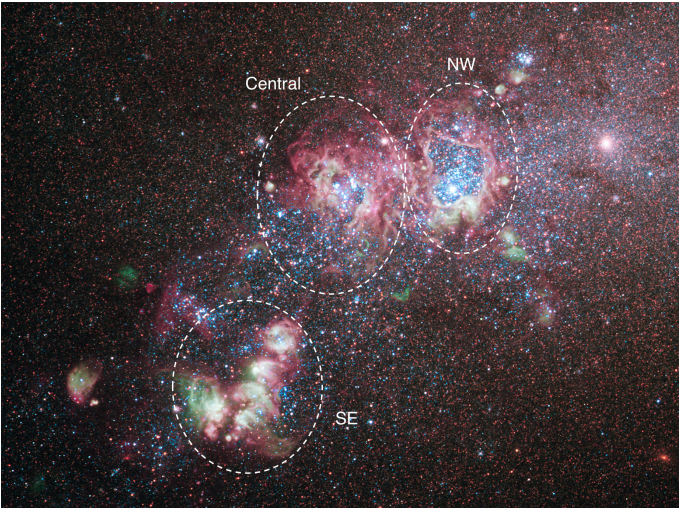}
\caption{A composite HST/WFC3 image of NGC 4214 from GO-11360 (PI: O'Connell). This image is a combination of red (F657N and F814W), green (F502N and F547M), and blue (F225W, F336W, F438W, and F487N) images. The morphology reveals many blue stars surrounded by expanding superbubbles. White circles mark three prominent star-forming complexes labeled as SE, Central, and NW complexes. Credit: NASA, ESA, and the Hubble Heritage (STScI/AURA)-ESA/Hubble Collaboration.}
\label{n4214comp}
\end{figure}

\begin{figure*}
\centering
\includegraphics[trim=3cm 0cm 0cm 0cm, clip=True, width=20cm]{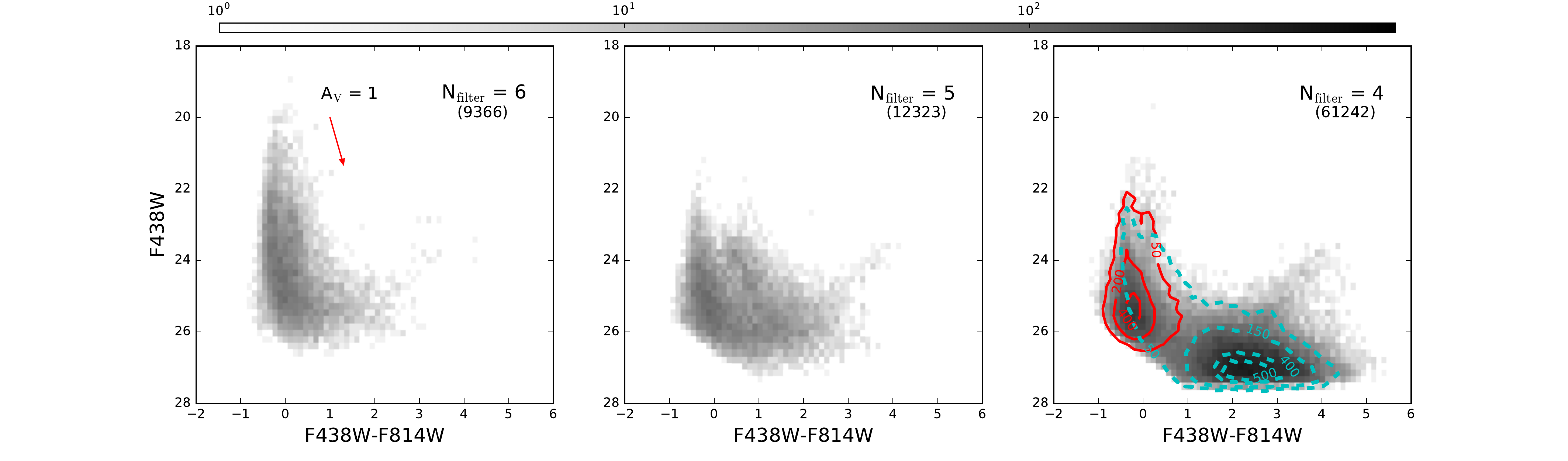}
\caption{In this study, we analyze a total of 82,931 stars that are detected at least in 4 bands. We present the optical CMDs of stars that are detected in all six bands (left; 9,366 stars), any five bands (middle; 12,323 stars), and any four bands (right; 61,242 stars). In the right panel, we also show the distribution of stars with 2 NUV and 2 optical detections (red solid contours). These stars occupy the blue (F438W--F814W $<$ 1~mag) part the CMD. On the other hand, stars with 2 optical and 2 NIR detections (dashed blue contours) dominate the red (F438W--F814W $>$ 1~mag) part of the CMD. An arrow indicates the direction of the reddening vector 
\label{nfilter}}
\end{figure*}

There are three prominent star-forming complexes in NGC 4214 (SE, Central, and NW complexes as roughly marked in Figure~\ref{n4214comp}), within which we expect most ionizing radiation to be produced. Enhanced star formation in these regions started $\sim$30~Myr ago \citep{williams11}, and the galaxy has strong Wolf-Rayet features in some of its HII region spectra \citep{kobulnicky96}, consistent with its high recent SFR. NGC 4214's total molecular gas mass is M$_{\rm H2} =$ 5.1$\times$10$^{6}$~\msun\,\citep{walter01}, the total atomic gas mass is M$_{\rm HI} =$ 4.1$\times$10$^{8}$~\msun\,\citep{walter08}, the total dust mass is M$_{\rm Dust} =$ 1.62$\times$10$^{6}$~\msun\,\citep{hermelo13}, making it a gas-rich dwarf galaxy. Its metallicity is low \citep[$\sim$0.25~\zsun;][]{kobulnicky96}, similar to that of Small Magellanic Cloud. Its low metallicity results in the low dust content in NGC 4214. \citet{hermelo13} reported a gas-to-dust mass ratio of 350--470, which is 3.5--4.7 times higher than the Milky Way average value. Due to its proximity \citep[3.04$\pm$0.05~Mpc;][]{dalcanton09}, NGC 4214 has been mapped in many wavelengths from the IR (e.g., with Spitzer and Herschel), through the optical (e.g., with HST), UV (e.g., with GALEX), and X-ray (Chandra). This large set of ancillary data for NGC 4214 enables one to probe its detailed ISM properties \citep[e.g.,][]{mackenty00, ubeda07a, cormier10, hermelo13} and its star formation history (SFH) using resolved stellar populations \citep[e.g.,][]{mcquinn10, williams11, weisz11}.

\section{Data and Photometry} \label{sec:data}

\subsection{Simultaneous Six-filter Photometry}
We use {\it HST} images obtained with the WFC3/UVIS (F225W, F336W, F438W, and F814W bands) and WFC3/IR (F110W and F160W bands) cameras under the program GO-11360 (PI: O'Connell). The exposure times for F225W, F336W, F438W, F814W, F110W, and F160W are 1665s, 1683s, 1530s, 1339s, 1198s, and 2397s, respectively. The data have sufficient depth for our science goals, which focus on luminous O/B stars. Figure~\ref{n4214comp} shows a color image of NGC 4214 created from multi-wavelength HST/WFC3 data. 

The resolved stellar photometry is derived as follows. First, cosmic rays (CRs) are identified in the WFC3/UVIS data by first running the IDL routine {\tt lacosmic} \citep{vandokkum01} on all of the UVIS data. Further CRs are identified by running {\tt astrodrizzle} \citep{gonzaga12} on data from each band with exposure times larger than 50 seconds.

Once the CRs are identified and flagged in the data quality (DQ) extensions of the calibrated (``FLT'') images, each image is processed through the DOLPHOT \citep{dolphin00} task {\tt acsmask} or {\tt wfc3mask} as appropriate. This task applies the DQ flags as well as the pixel area map to the FLT science (SCI) extensions in order to mask out bad pixels and CRs, and then calibrates the flux in each pixel by the area coverage on the sky. The FLT images were then split into their individual CCDs using the DOLPHOT task {\tt splitgroups}.

\begin{figure}
\includegraphics[trim=2.5cm 0cm 1.5cm 0cm, clip=true, width=9cm]{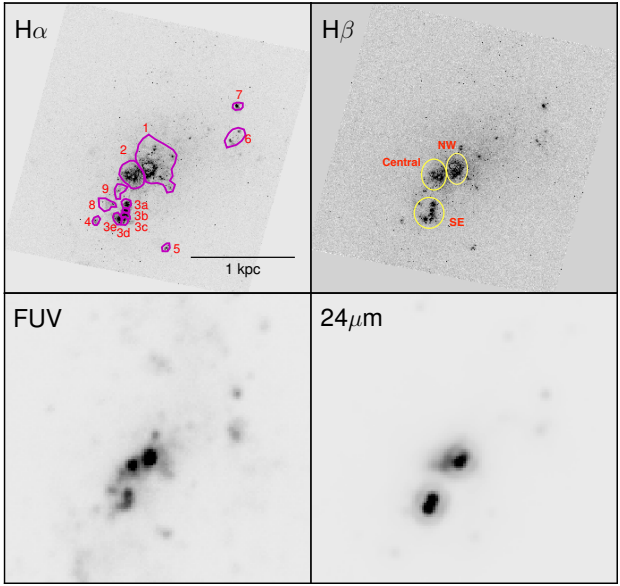}
\caption{Ancillary Data for NGC 4214. From top left to bottom right, the narrow-band, pre-continuum removal images of H$\alpha$ (HST/WFC3 F657N) and H$\beta$ (HST/WFC3 F487N), Spitzer 24$\mu$m image, and GALEX FUV image. Most emission associated with SF is confined in this field of view. Irregular polygons in the top left panel denote the 13 prominent SF regions defined based on the H$\alpha$ brightness and morphology. We measure the local escape fractions in these 13 SF regions individually in Section~\ref{sec:localEsc}.
\label{ancillary}}
\end{figure}

To simultaneously find and measure stars in all six bands, we first align all data to a relative precision of $\sim$0.2 pixel (8 milli-arcseconds). We measure the positions of stars in each CCD of each exposure to provide the basis for astrometric alignment. These single CCD DOLPHOT catalogs are culled for artifacts using the same criteria as those used for the single camera photometry in \citet{dalcanton12}. Once we produce catalogs for each individual CCD of each exposure, we use astrometry.net \citep{lang10} to find the best alignment of all catalogs. This software generates header update scripts that we applied to the original FLC fits images from MAST to produce astrometrically-aligned images from all visits in all bands.

With the original FLT fits images updated with the precision-aligned headers, we run them all through the DOLPHOT preprocessing ({\tt lacosmic}, {\tt astrodrizzle}, {\tt wfc3mask}, {\tt calcsky}), to prepare to run the final photometry. We use the F438W {\tt astrodrizzle} output as the DOLPHOT reference image for astrometry. Our final alignment to the F438W astrodrizzled reference image, as measured by DOLPHOT, has an root-mean-square scatter of 0.3 pixels or better in all of the IR exposures and 0.2 pixels or better in all of the UVIS exposures.

The output of DOLPHOT includes all $>$ 3.5$\sigma$ peaks in the full data set. It therefore accounts for flux from neighbors as precisely as possible, but also contains many measurements that are due to artifacts (warm pixels, poorly-masked CRs) or noise. We therefore cull the catalogs based on quality parameters measured by DOLPHOT. Specifically, in each band, we only accept measurements with a signal-to-noise $>$ 4, a square of sharpness $<$ 0.15, and crowding $<$ 1.3 (2.25) for the WFC3/UVIS (WFC3/IR). These cuts produce sharp-featured CMDs of accepted stars. We refer the reader to \citet{williams14} for details of comparable simultaneous 6-filter photometry for the PHAT \citep[Panchromatic Hubble Andromeda Treasury,][]{dalcanton12} data set. In our final catalog, there are $\sim$83,000 resolved stars that pass the above quality cuts in \textit{at least four bands} out of six. This is our 4-band detection requirement, which enables much more reliable SED fitting by ensuring the combination of the UV, optical, and IR to resolve the degeneracy between the dust and stellar temperature \citep[see][]{gordon16}. We note that we still utilize negative fluxes for one or two bands where stars were not detected, because these negative fluxes provide upper limits for those bands when fitting SEDs.

Figure~\ref{nfilter} presents the optical CMDs of stars that are detected in all six bands (left; 9,366 stars), only five bands (middle; 12,323 stars), and only four bands (right; 61,242 stars), yielding a total of 82,931 stars. Stars that are detected in all six bands are mostly blue and bright, whereas stars that are detected only in any four bands are fainter and redder. Even for the blue (F438W--F814W $<$ 1~mag) stars, the brightest stars with four-band detections are $\sim$2~mag fainter than those with six-band detections. Stars with five-band detections occupy the middle position on the CMD. In the CMD of four-band detection stars, there are two distribution concentrations of stars: one at F438W--F814W $<$ 1~mag, and the other at F438W--F814W $>$ 1~mag. The bluer cluster consists of stars with four-band detections in the UV and the optical filters (red solid contours), while the majority of stars in the redder cluster are detected in the optical and the IR filters (cyan dashed contours).

\subsection{Ancillary Data} 
In addition to the HST/WFC3 broad-band imaging data, we use the WFC3/UVIS narrow-band images of H$\alpha$ (F657N) and H$\beta$ (F487N) lines and a medium-band image in F547M filter (also from GO-11360; PI: O'Connell) to derive an extinction-corrected H$\alpha$ emission-line map. We will describe how to derive the extinction-corrected H$\alpha$ emission-line map in detail in Section~\ref{calc_ngas}. Briefly, we first subtract scaled continuum images from narrow-band H$\alpha$ and H$\beta$ images. The subtraction removes the contribution from stellar photospheres and leaves only the nebular emission. We also account for the spatially-varying [NII] contribution to the F657N image using the measured correlation between the [NII]/H$\alpha$ ratio and the H$\alpha$ intensity \citep{maiz-apellaniz98}. We then use the ratio map of continuum-subtracted H$\alpha$/H$\beta$ to calculate an extinction map using the Balmer decrement assuming Case B recombination, and correct the H$\alpha$ emission-line map for extinction by taking the effect of the gas/dust geometry into account. The extinction-corrected H$\alpha$ map is used to estimate the number of ionizing photons consumed by photoionization in the surrounding ISM (see Section~\ref{sec:fesc}). We also use GALEX FUV (GI4-095, PI: Lee) and Spitzer 24$\mu$m \citep{dale09} images to briefly compare with the predicted FUV map and $A_V $ map derived from stellar SED fitting. Figure~\ref{ancillary} shows the H$\alpha$ (F657N), H$\beta$ (F487N), and GALEX FUV images.

\section{The BEAST Settings for NGC 4214}\label{n4214_setting}
The BEAST is a probabilistic fitting tool, and it fits libraries of stellar model SEDs with dust extinction to the observed broad-band SEDs of individual stars. We can, therefore, fit multi-wavelength photometry of individual stars simultaneously for both the stellar properties (e.g., age, mass, metallicity) and the intervening dust properties (column density, composition, and grain size distribution). It is possible to constrain both the stellar and dust extinction parameters because the shape of stellar SED covering the UV to IR responds to dust and stellar temperature in different ways, with the UV filters, in particular, being essential to breaking the degeneracy between dust extinction and stellar temperature. 

Therefore, by running the BEAST on multi-wavelength photometry of resolved stellar populations, we can infer the intrinsic emission for each bright star in any bandpass, allowing us to create a high resolution map of the intrinsic ionizing photon production rate. In this section, we describe how we derive the intrinsic properties of individual stars and intervening dust from the stellar SED fitting technique in more detail. 

The BEAST is optimized for large multi-wavelength surveys of resolved stellar populations. A full detailed technical description of the BEAST can be found in \citet{gordon16}. They also showed that the application of the BEAST to the PHAT survey dataset \citep{dalcanton12, williams14} is successful, in that the results yield a reasonable Hertzsprung-Russell (HR) diagram, understandable variations in fit parameter precision, and a spatial map of the average dust column that correlates well with existing infrared derived dust surface densities \citep{draine14}. While the BEAST was originally developed for the PHAT survey, it is applicable to any multi-wavelength resolved star survey, including the NGC 4214 dataset, with appropriate priors. 

In the following subsections, we discuss applying the BEAST to NGC 4214; NGC 4214 has very different physical conditions from M31, and thus the proper priors on the stellar, dust, and noise models should be different from those presented in \citet{gordon16} for M31. We also modify the default BEAST dust model in the Far-UV (FUV) and Extreme-UV (EUV) regimes for this study. 

\begin{figure}
\includegraphics[width=11cm, trim=3cm 1.5cm 0cm 2cm, clip=true]{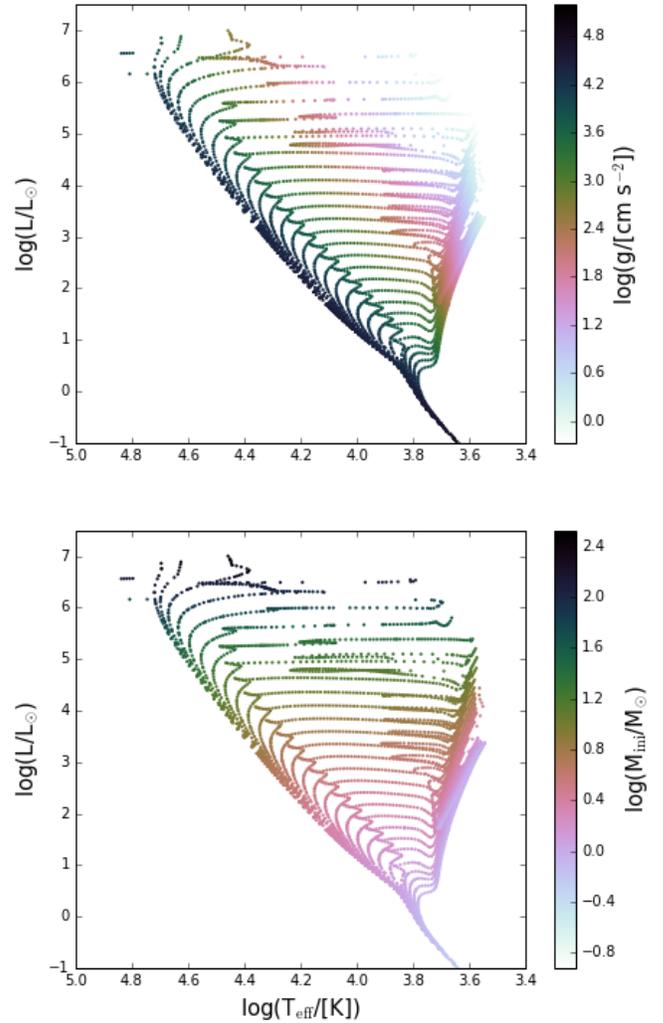}
\caption{The entire coverage of the stellar model grid is shown in a Hertzsprung-Russell diagram color-coded by $\log(g)$ and $\log(M_{\rm ini})$ in the top and bottom panels, respectively. When performing fitting, we trim this big grid to exclude models with fluxes 3$\sigma$ fainter than the faintest stars in all six bands. 
\label{params}}
\end{figure}

\begin{figure}
\includegraphics[width=9cm]{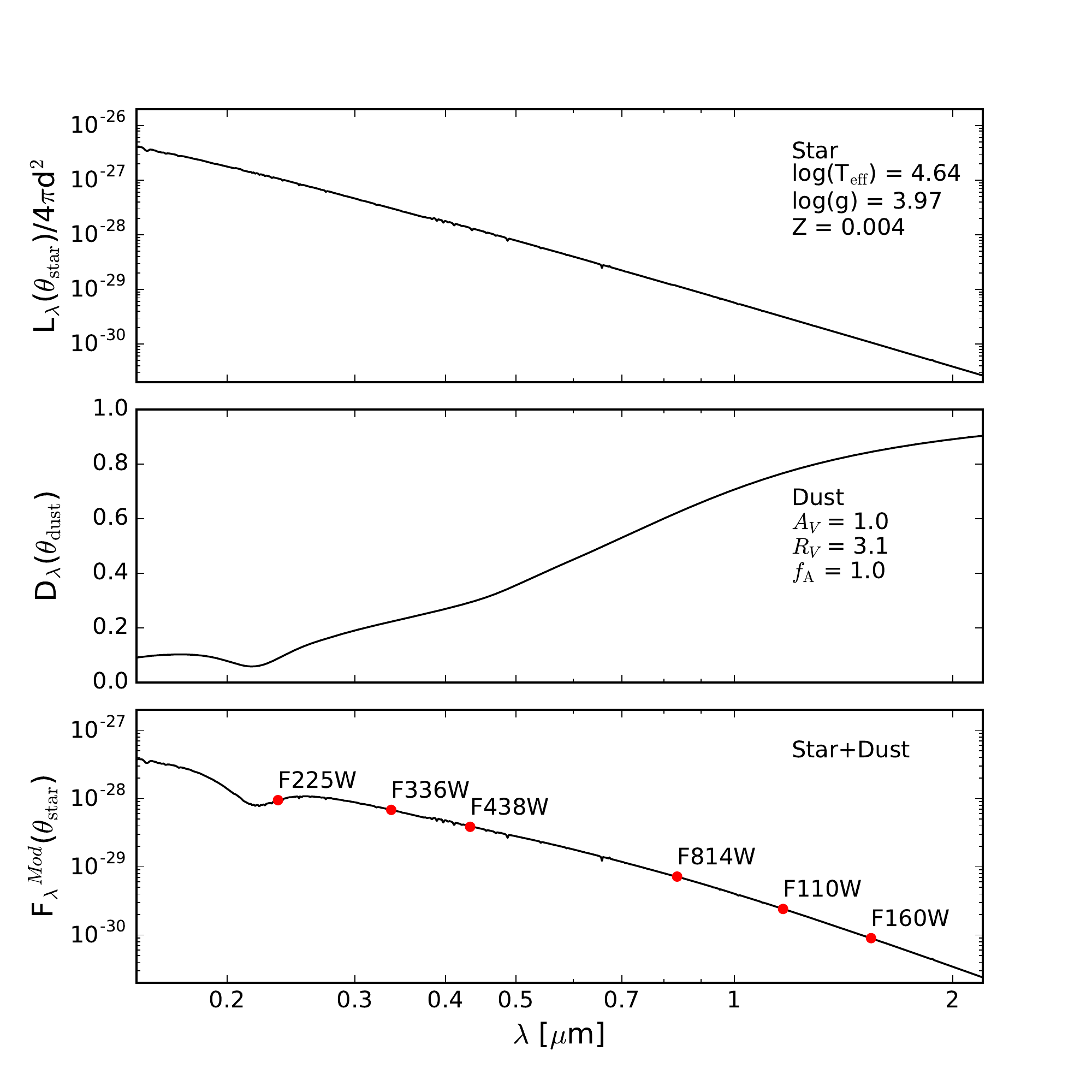}
\caption{The computation of a BEAST model SED for a dust extinguished star is shown graphically. The top panel gives the intrinsic stellar spectrum at the distance to NGC 4214, the middle panel shows the extinction by dust, and the bottom panel plots the full extinguished stellar spectrum. As an example, we show the integrated SEDs for the 6-filter bandpasses (solid circles) in the bottom panel.}
\label{fig_exp_spec}
\end{figure}

\subsection{Stellar Model} \label{beast_stellarmodel}
In this study, we use the PARSEC version 1.2S \citep{tang14} that fully describes massive stars (up to 336~\msun) for low metallicity (0.001 $\leqslant Z \leqslant$ 0.004). For the stellar atmosphere model, we use a merger between the local thermal equilibrium (LTE) grid from \citet{castelli04} and the non-LTE TLusty OSTAR and BSTAR grids from \citet{lanz03,lanz07}. In the overlap regions, the non-LTE model takes priority over the LTE model. As discussed in \citet{gordon16}, we construct the stellar model grid by mapping the stellar evolutionary model (defined by stellar initial mass $M_{\rm ini}$, stellar age $t$, and metallicity $Z$) onto stellar atmospheric model (defined by stellar effective temperature $T_{\mathrm{eff}}$, surface gravity $\log(g)$, and metallicity $Z$). Thus, ($M_{\rm ini},t,Z$) is our primary set of stellar parameters, denoted $\theta_{\mathrm{star}} = \{M_{\rm ini},t,Z\}$. The full coverage of our stellar model on the parameter space ($\log(T)$, $\log(L)$, $\log(g)$, and $\log(M_{\rm ini}$)) is shown in Figure~\ref{params}. To reduce computing time, we exclude unnecessary models whose fluxes are 3$\sigma$ below the survey sensitivity as measured from the AST results.

\subsection{Dust Model} 
A successful model of an observed stellar SED requires modeling emission spectra of stars as well as absorption spectra of the intervening dust. Emergent light from a star is extinguished by interstellar dust before reaching the observer. The degree to which the emergent light is extinguished is a strong function of wavelength. Extinction curves from the FUV to the NIR have been measured along many sightlines in the Milky Way (MW) \citep{cardelli89, fitzpatrick99, valencic04, gordon09}, as well as in the Magellanic Clouds \citep{gordon98, misselt99, maizappellaniz12} and M31 \citep{bianchi96, clayton15}. Most sight-lines in the Magellanic Clouds behave very differently than the MW $R_{\rm V}$-dependent relationship \citep{gordon03}. For example, the extinction curves in the Small Magellanic Cloud (SMC) star-forming bar show nearly linear dependence on $\lambda^{-1}$. Also, there is no strong 2175~\AA\ extinction bump in the extinction curves of the SMC star-forming bar. 

\begin{figure}
\includegraphics[width=9cm, trim=1cm 0.5cm 1cm 0.5cm, clip=true]{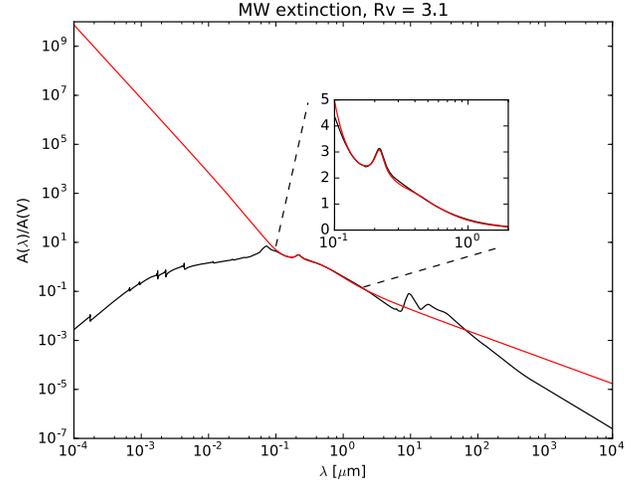}
\caption{Below the Lyman limit, the behavior of the BEAST default extinction curve (red solid), which is based on observations at wavelengths longer than 1150~\AA\,(\textit{IUE} satellite cutoff), is obviously different from the theoretical extinction curve (black solid). The BEAST simply extrapolates the default extinction curves at wavelengths shorter than $\lambda$ of 1150~\AA. This results in an unphysical rise in the extinction curves towards the LyC regime. Thus, we modify the default BEAST curves to adopt the theoretical curves below 912~\AA. For a smooth transition between the BEAST and theoretical curves in the FUV regime, we perform the weighted average at the wavelength range between 912~\AA\, $ < \lambda <$ 1695~\AA. The inset magnifies the FUV to NIR regime. 
\label{MW_ext_draine}}
\end{figure}

It has been found that the observed extinction curves in the Local Group lie between the average MW extinction curve \citep[e.g.,][]{cardelli89, fitzpatrick99, gordon09} and the average SMC star-forming bar extinction curve \citep{gordon03}. Thus, \citet{gordon16} introduced a mixture model for the dust extinction that effectively interpolates between a MW extinction model (``$\mathcal{A}$'' component) and a SMC Bar extinction model (``$\mathcal{B}$'' component). The primary set for dust model parameters is $\theta_{\mathrm{dust}} = \{A_V, R_V, f_\mathcal{A} \}$, where $f_\mathcal{A}$ is the fraction of the $\mathcal{A}$-type extinction. The default mixture model utilizes \citet{fitzpatrick99} for the $\mathcal{A}$ component and \citet{gordon03} for the $\mathcal{B}$ component. Both studies derived UV extinction curves based on the International Ultraviolet Explorer (IUE) spectra, which has a wavelength cutoff at 1150~\AA.

Figure~\ref{fig_exp_spec} gives a graphical representation of how the BEAST computes a model SED by showing its intrinsic spectrum, the effects of dust, the extinguished spectrum, and the band integrated SED. The fundamental principle of the SED fitting is to compare these model fluxes with photometric observations. However, the actual SED fitting process involves more complexities, which we discuss in the next two subsections.

Although the BEAST's dust mixture model can describe the full range of observed extinction curves in the Local Group from FUV ($\lambda >$ 1150~\AA) to NIR, inferring the absorption of LyC photons requires knowledge of the extinction curves in the extreme UV below the Lyman limit, where there is currently no observational data. Simply extrapolating the default extinction curve to short wavelengths produces an unphysically steep rise in extinction below the Lyman limit. Instead, we adopt theoretical extinction curves below the Lyman limit \citep{weingartner01, draine03}. The model extinction curves increase smoothly as wavelength decreases as seen in the observed extinction curves \citep[e.g.,][]{gordon09}, peak at around 750~\AA\, and then decrease. This carbonaceous-silicate grain model successfully reproduces the observed interstellar extinction, scattering, and NIR emission \citep{li01}. 

Figure~\ref{MW_ext_draine} presents an example showing how we extend extinction curves down to the LyC regime. To merge the theoretical curves onto the BEAST default extinction curves (i.e., the empirical curves) between 912~\AA\, $< \lambda <$ 1695~\AA, we compute a weighted mean between the theoretical and the BEAST curves for a smooth transition. Specifically, we first obtain the $R_{\rm V}$-dependent model extinction curves with 2 $\leq$ $R_{\rm V}$ $\leq$ 6 by extrapolating and interpolating the theoretical MW extinction curves provided for $R_{\rm V}$ values of 3.1, 4.0, and 5.5. The weight linearly varies on the BEAST curves from zero at 912~\AA\, to one at 1695~\AA. The opposite weight is applied to the theoretical extinction curves, and the resulting weighted average produces a smooth transition from the BEAST to the model curves. 

Thus, our final extinction curves have three regimes: (1) the default BEAST curve regime ($\lambda >$ 1695~\AA), (2) the transition regime from the BEAST curve to the model curve (912~\AA\, $ < \lambda <$ 1695~\AA), and (3) the model curve regime ($\lambda <$ 912~\AA\,). We repeat the same process for the SMC star-forming Bar extinction curve with a fixed $R_{\rm V}$ of 2.74 as well. The theoretical curve for the SMC star-forming bar extends only down to 100~\AA\, while those for the MW extend to 1~\AA. In this work, we integrate spectra over 228~\AA\, < $\lambda$ < 912~\AA\, to calculate the number of hydrogen ionizing photons. This is because the contribution of ionizing photons with $\lambda <$ 228~\AA, corresponding to He$\,\textsc{ii}$ ionization energy of $\sim$54.42~eV (4~Rydberg), to the photoionization of hydrogen is negligible due to its small photoionization cross-section.  

\begin{figure}
\includegraphics[width=11cm, trim=3cm 14cm 0cm 2cm, clip=true]{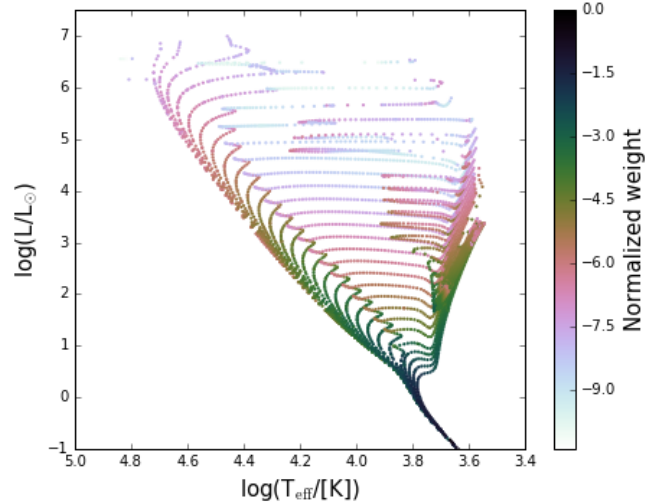}
\caption{Mapping of the stellar mass and age priors into the log($L$) versus log($T_{\mathrm{eff}}$). Darker (lighter) colors denote higher (lower) prior probabilities. As expected, higher weights are assigned to low-mass main-sequence stars and older red-giant branch stars.}
\label{prior}
\end{figure}

\begin{deluxetable*}{lcc}
\tablecaption{Photometry and BEAST Results \label{finCAT}} 
\tablewidth{0pt}
\tablehead{
\colhead{Column names} &
\colhead{Description} & \colhead{Units} }
\startdata
RA               & Right ascension (J2000)                                     & deg      \\
DEC            &  Declination (J2000)                                            & deg      \\ 
F225W$\_$VEGA                   & F225W Vega magnitude          & mag       \\
F336W$\_$VEGA                   &  F336W Vega magnitude         & mag      \\ 
F438W$\_$VEGA                   &  F438W Vega magnitude        & mag      \\
F814W$\_$VEGA                   &  F814W Vega magnitude        & mag     \\
F110W$\_$VEGA                   &  F110W Vega magnitude         & mag         \\
F160W$\_$VEGA                   &  F160W Vega magnitude         & mag        \\
logT$\_$[p16,p50,p84]    &  log of effective stellar temperature          & K      \\
logA$\_$[p16,p50,p84]    & log of stellar age                                      &  yr \\
logg$\_$[p16,p50,p84]    & log of stellar surface gravity                     & cm~s$^{-2}$    \\
M$\_$ini$\_$[p16,p50,p84,Exp]    & initial stellar mass                             & \msun    \\
Av$\_$[p16,p50,p84,Exp]  & internal extinction in the V-band           & mag   \\
logNION$\_$int$\_$[p16,p50,p84]  & log of $\dot{N}_{\rm int}$    &  \# of LyC photons per second     \\
 \enddata
\tablecomments{Six-filter photometry and the BEAST fits used in this study for 82,931 stars.}
\end{deluxetable*}

\subsection{Priors}\label{beast_prior}
The BEAST SED fitting includes the options of using prior distributions on the model parameters. When parameters are well understood, we specify physically motivated priors to improve the statistical accuracy of the fitting, particularly in the limit of noisy data. Relatively well-established knowledge about stars and dust in a galaxy from previous studies generally serve as a good approximation to what our posterior beliefs should be. For example, it is known that a stellar initial mass function (IMF) favors the production of low mass stars \citep[see the review of][]{Bastian10}. To reflect the expected distribution of real stars, we can select proper priors on stellar mass and age to be consistent with a given initial mass function and SFH, and apply those priors to the model grid points by assigning corresponding weights. 

Figure~\ref{prior} shows an example of mapping of the stellar mass and age priors into the model log($T_{\mathrm{eff}}$)-log($L$) space. The BEAST adopts a flat prior on $A_{\rm V}$ and on the 2D space of $R_{\rm V}$ versus $f_\mathcal{A}$ within the limited range of $R_{\rm V}$ between 2 and 6 \citep[see Fig. 7 in][]{gordon16}. In this study, we adopt flat priors on $A_V =$ between 0.0 and 8.0. We set the maximum $A_V$ to be 8~mag because the depth of our data would not allow us to recover a star with dust extinction higher than this limit at all, even for the brightest star in our sample. Note that the resulting posterior distribution follows the prior distributions in case of low signal-to-noise measurements. 

In addition to the BEAST's default/standard priors, we put much stronger priors on a number of parameters that are suitable for NGC 4214. These include the distance to NGC 4214 \citep[3.04~Mpc;][]{dalcanton09}, metallicity \citep[$Z =$ 0.004, which appears not to vary across the galaxy significantly;][]{kobulnicky96}, and the MW foreground extinction \citep[$A_{\rm V}$ = 0.06~mag;][]{schlafly11}. To count the effect only by the internal dust on $f_{\rm esc}$, we apply the MW foreground extinction to the stellar model grid before modeling the SEDs and apply the dust mixture model to measure the internal dust during the actual SED fitting.

\subsection{Noise Model}\label{sec:noise}
Another virtue of successful SED modeling includes appropriate modeling of observational effects. Observational uncertainties of stars in external galaxies result from a combination of photon noise and crowding. To properly account for these observational effects in our SED fitting, we construct a full observational ``noise model'' by performing extensive artificial star tests (ASTs). Specifically, we construct the full noise model for individual SED models by running extensive 6-filter ASTs. For each SED model, we insert an artificial star with a known SED into the observed image and re-run photometry in the same way we did for real stars. These ASTs capture complexities due to the non-linear interaction between the photon and crowding noise, allowing us to generate a noise model that describes the differences between the input and output fluxes in all filters and the covariances between the offsets in different filters. For further details, we refer the reader to \citet{gordon16}. 
 
The properties of the noise model depend strongly on stellar number density, given that crowding usually dominates photometric uncertainties particularly at longer wavelengths \citep[Figure 17 in][]{gordon16}. We therefore compute the noise model in bins of stellar number density, which we define as the number of stars per arcsec$^2$. This density is calculated based on 20 nearest neighbors using only stars with 20 $<$ F438W $<$ 24, for which detections are $\sim$100\% complete in F438W. We divide the resulting stellar number density map up into eight different stellar number density bins that are evenly spaced on a log scale. Then we calculate the noise model for each stellar number density bin. 

For each stellar number density bin, we sample the input artificial stars from the model SED grid to guarantee that the noise model is calculated for realistic populations of stars. First, we trim the model grid to only keep the models brighter than 3$\sigma$ below flux$_{\rm min,obs}$ in a given stellar number density bin. This choice removes models that are too faint to be detected in our observations, saving computational time, but still accounting for upscattered faint stars. We then select 630 unique models from the trimmed grid that covers the entire parameter space of interest uniformly, but coarsely (it is not computationally feasible to run ASTs over every model SED in each density bin). We require 20 independent realizations for each artificial star with random spatial locations in a given stellar number density region to draw a reasonably measured mean offset between the true and measured fluxes, i.e., the bias. We also take the full covariance and the model-independent absolute flux calibration into account. We then interpolate the resulting noise models on the full SED models. 

We repeat the same procedure for all eight stellar number density bins. The equal number of artificial stars in each density bin ensures the same quality of noise model at all stellar number density ranges. At the same time, it effectively allows more rigorous ASTs in higher density bins because the areal coverage for a higher density bin is smaller. In total, we use 100,800 ASTs ($=$ 8 stellar number density bins $\times$ 630 unique artificial stars $\times$ 20 realizations) to construct the full noise model for individual dust-reddened stellar models by running {\it simultaneous} six-filter ASTs. Note that both the numbers of unique artificial stars and independent realizations are empirically set by two factors: (1) obtaining a realistic and reasonable noise model, and (2) our limited computational resources.

\section{Results}\label{sec:results}

\begin{figure}
\centering
\includegraphics[trim=0cm 0cm 0.5cm 0cm, width=9cm, clip=True]{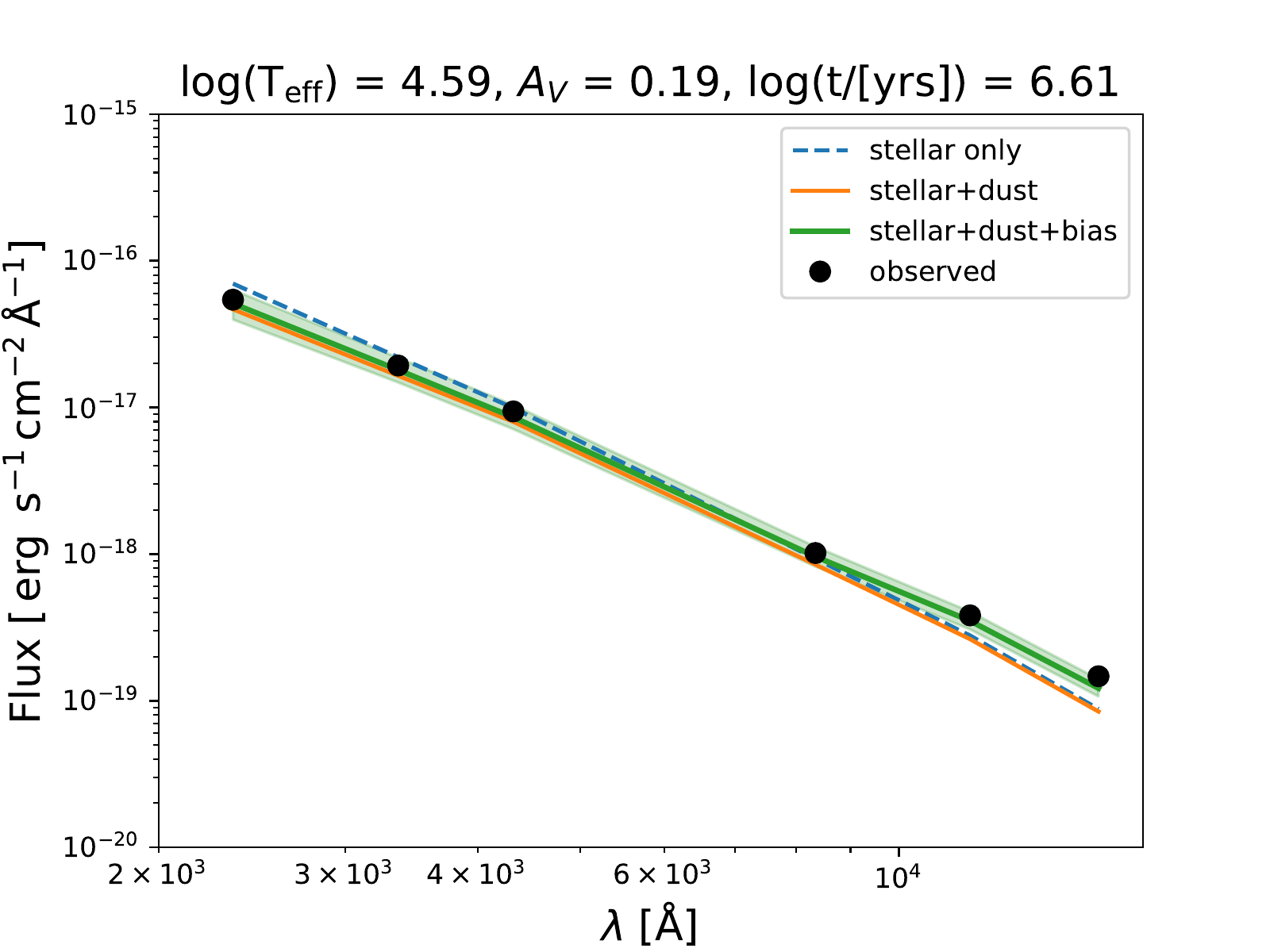}
\includegraphics[trim=1cm 0cm 0.5cm 1cm, width=9cm, clip=True]{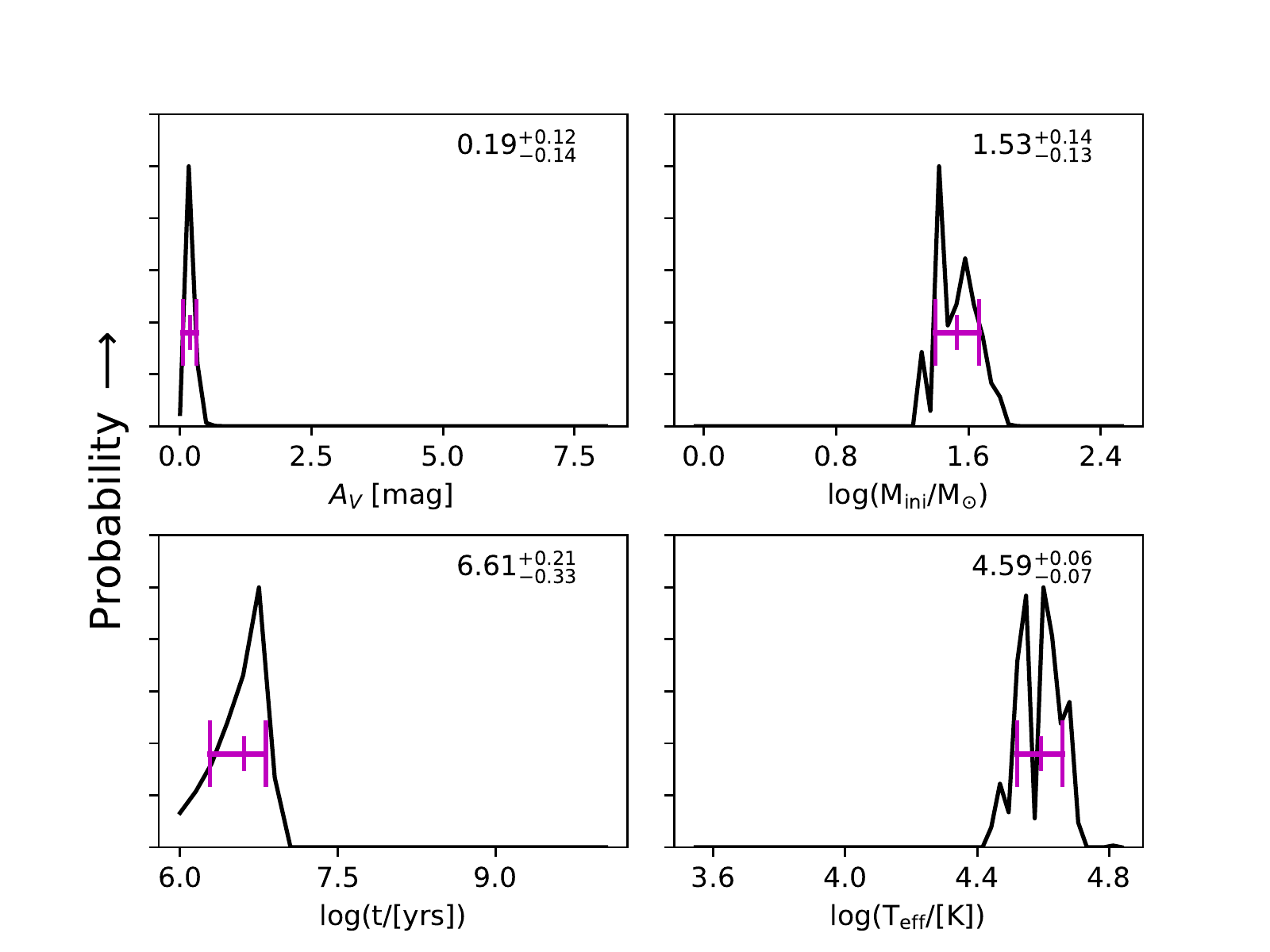}
\caption{{\it Top:} An example of the SED fitting results for a young MS star in NGC 4214. Black circles denote the observed six-filter photometry. The 50th percentile model (green solid line) is shown along with shaded colored regions indicating the range of models that fit within 1$\sigma$. We also present the the stellar model alone (blue dashed line) and the physical (star+dust) model alone (orange solid line). The impact of the bias in the photometric measurements is clearly seen in the IR and illustrates the importance of a full physical and observational model for this type of fitting. {\it Bottom:} The 1D pPDFs for $A_{\rm V}$, log(M$_{\rm ini}$), log($t$), and log($T_\mathrm{eff}$) parameters. The 50\%$\pm$33\% values are given numerically on the upper right corner as well as graphically (magenta lines).  
\label{exampleSED_hot}}
\end{figure}

\begin{figure*}
\centering
\includegraphics[width=18cm, trim=2.5cm 0cm 2cm 1cm, clip=true]{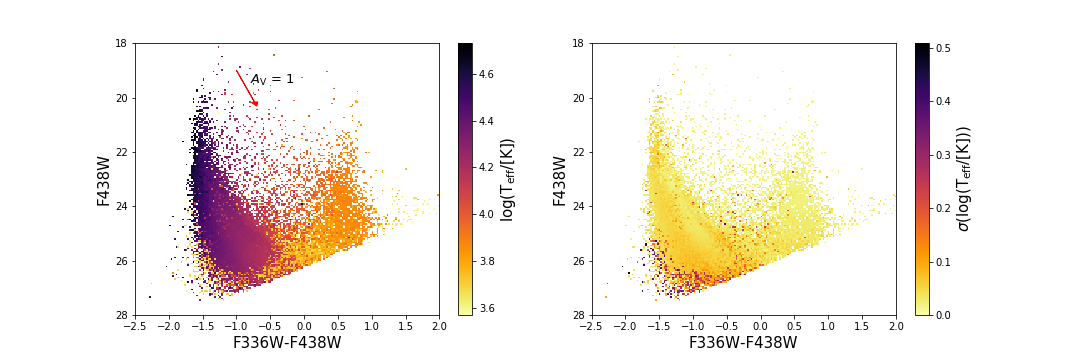}
\includegraphics[width=18cm, trim=2.5cm 0cm 2cm 1cm, clip=true]{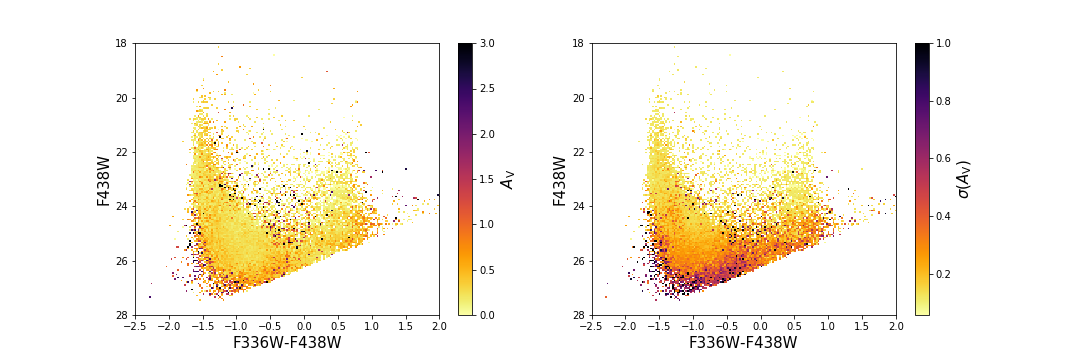}
\caption{{\it Top:} The CMD color-coded by the 50th percentile values for log($T_{\rm eff}$/[K]) (left), and by 1$\sigma$ uncertainty on log($T_{\rm eff}$/[K]) (right). {\it Bottom:} The CMD color-coded by the 50th percentile values for $A_{\rm V}$ (left), and by 1$\sigma$ uncertainty on $A_{\rm V}$ (right).  
\label{cmdTAv}}
\end{figure*}

Figure~\ref{exampleSED_hot} shows an example of the BEAST SED fitting for a massive young star in NGC 4214. In the top panel, solid circles are observations and the colored lines show the 50th percentile models; blue dashed line, orange solid line, and green thick line represent the stellar model, the star+dust model, and the star+dust model including the observational bias, respectively. The green shaded envelope indicates the 68\% confidence interval of the full model (i.e., star+dust+bias). The impact of the bias on the photometric measurements is seen in the IR and illustrates the importance of a full physical and observational model for this type of fitting.

In the bottom panels, we show the 1D posterior probability distribution functions (pPDFs) for the key parameters of this study -- $A_{\rm V}$, log(M$_{\rm ini}$), log($t$), and log($T_\mathrm{eff}$) parameters. This star is fit as a massive young star without any confusion in SED fitting. The 50\%$\pm$33\% values of each parameter are given numerically as well as graphically (magenta lines) in each panel. In the following figures and analysis, the reported 1$\sigma$ uncertainties for each parameter are estimated from the 16th and 84th percentile values. The full catalog is available in FITS format from MAST\footnote{https://archive.stsci.edu/prepds/uvescape/}. Table~\ref{finCAT} describes the columns in the catalog.

\begin{figure*}
\centering
\includegraphics[width=18cm, trim=2cm 3cm 3cm 3cm, clip=True]{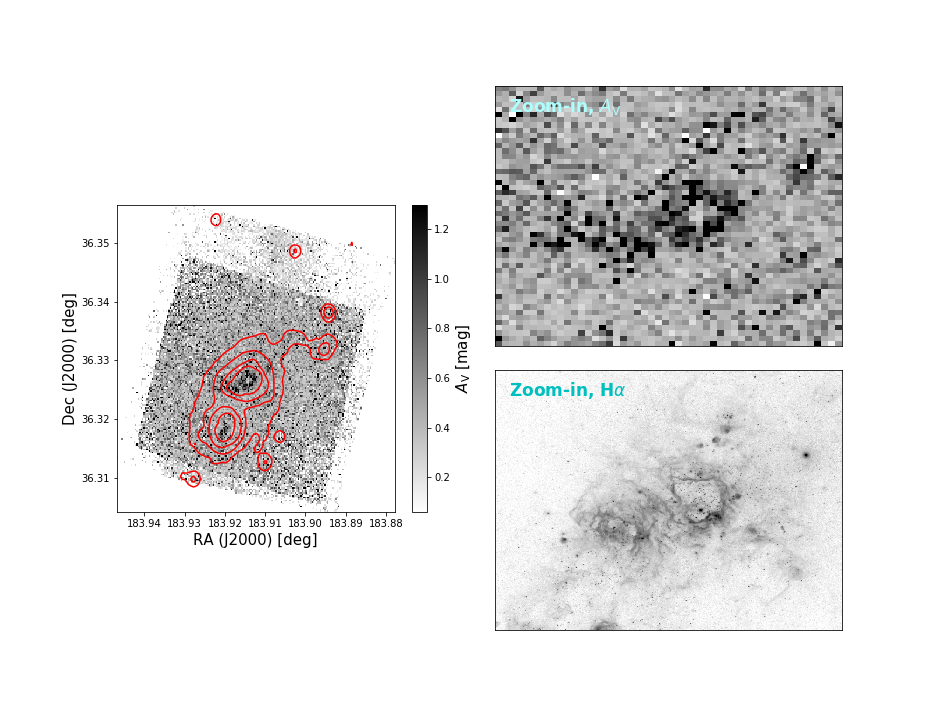}
\caption{{\it Left:} Comparison between the derived $A_{\rm V}$ map (gray scales) and the Spitzer MIPS 24$\mu$m emission (red contours). They show a good spatial correlation in that higher $A_{\rm V}$ values cluster near the regions where 24$\mu$m emission is stronger. {\it Top right:} A zoom-in around the Central and NW SF complexes. Our SED modeling reproduces the presence of dusty shells around these SF complexes {\it Bottom right:} H$\alpha$ image cutout around the same SF complexes for comparison. 
\label{avMaps}}
\end{figure*}

\subsection{$T_{\rm eff}$ and $A_{\rm V}$}
In this subsection, we discuss the two key parameters, $T_{\rm eff}$ and $A_{\rm V}$, that are recovered from our SED fitting. Among the stellar parameters mentioned in Section~\ref{beast_stellarmodel}, the stellar effective temperature is most closely tied to the ionizing photon production rate. While there is a trend of the ionizing photon production rate with the stellar age for young massive stars, the scatter is relatively large due to the inherent difficulty in the age determination of individual zero-age MS stars (particularly at age $<$ 20~Myr) based on their photometric properties \citep[e.g.,][and references therein]{soderblom14}. Thus, we employ the effective temperature as the primary tool to select ionizing stars in the following analysis. Also, the derived dust properties along the line of sight to individual stars allow us to properly infer the amount of intrinsic ionizing photons from individual stars.

The left two panels in Figure~\ref{cmdTAv} show color-magnitude diagrams (CMDs) color-coded by the 50th percentile values for log($T_{\rm eff}$) and $A_{\rm V}$. Unsurprisingly, blue main-sequence (MS) stars have high effective temperature while faint MS stars and red giant branch (RGB) stars have lower effective temperature. About 87\% of stars in our sample show low dust extinction ($A_{\rm V} <$ 1~mag), as expected for a low-metallicity system. However, there are some intrinsically hot, but obscured stars, which would be expected for young stars that are still (partially) embedded in their nascent molecular clouds. These stars move along the reddening vector plotted on the CMD. If they were highly reddened hot stars, they might become too faint to meet our 4-band detection requirement; we will discuss the completeness of our data for hot stars further in Section~\ref{fuv_lyc}.

The right two panels in Figure~\ref{cmdTAv} show the same CMDs with 1$\sigma$ uncertainties for each parameter. Uncertainty in log($T_{\rm eff}$) is small ($<$ 0.1~dex) at most temperatures, except for the stars at the bottom of the blue MS. Their derived temperatures range between 3.78 $\lesssim$ log($T_{\rm eff}$) $\lesssim$ 4.1. In this temperature range, modeling broad-band SEDs suffers most from the intrinsic degeneracy between the reddening and stellar temperature due to the non-monotonic behavior in the Balmer jump strength \citep[see Figs. 6 \& 22 in][]{romaniello02}. However, the lowest temperature of interest for this study is $\sim$25,000~K (log($T_{\rm eff}$) $\simeq$ 4.4), corresponding to early B spectral type stars, and thus well outside the range of stellar temperature with significant uncertainties. Small uncertainties in high log($T_{\rm eff}$) secure reliable measurement of the intrinsic ionizing photon production rate.

$A_{\rm V}$ is less well constrained than log($T_{\rm eff}$). In addition to the parameter determinations' vulnerability to the dust/temperature degeneracy, fainter stars also tend to have higher uncertainties in $A_{\rm V}$ due to lower signal-to-noise ratio. Among cool stars with log($T_{\rm eff}$) $<$ 3.78 (i.e., cooler than F-type), 91\% of them show high fractional uncertainty in $A_{\rm V}$ ($>$ 50\%), while they still have low absolute $A_{\rm V}$ values and relatively small uncertainties in log($T_{\rm eff}$). We find that most of these cool stars are not detected in UV filters, making it hard to constrain their dust column density precisely; however, these stars also contribute negligible ionizing flux. Contrary to cool stars, stars hotter than early B-type have lower uncertainties in $A_{\rm V}$; roughly 40\% of hot stars have fractional uncertainties in $A_{\rm V}$ less than 50\%. This is because most of them are detected in 5 or 6 filters covering from NUV to NIR, thus significantly improving the precision and accuracy in the SED modeling by effectively breaking degeneracy between dust and stellar effective temperature.

In Figure~\ref{avMaps}, we compare the spatial distribution of the total (i.e., MW foreground $+$ internal dust) line-of-sight dust $A_{\rm V}$ (gray scale) to Spitzer MIPS 24~$\mu$m emission (red contours), which mainly arises from dust heated by young stellar populations. On the derived $A_{\rm V}$ map, which represents median 50th percentile values in each 18~pc$\times$14~pc spatial bin, one can see imprints of two distinct footprints; a smaller footprint for WFC3/IR and a larger footprint for WFC3/UVIS. The majority of stars in our catalog are cool and/or faint stars, and thus they are mostly confined to the smaller WFC3/IR footprint. Stars in the WFC3/IR footprint are detected in 4 or more bands with any combination of UV, optical, and IR. However, there are some stars that are only associated with the larger WFC3/UVIS footprint (i.e., outside the WFC3/IR footprint). These stars have to be detected in all F225W, F336W, F438W, and F814W to fulfill the $\geq$ 4-band detections criterion, leading to a bias towards UV-bright stars. Indeed, these stars are identified as more massive and hotter stars in stellar mass/temperature maps, discussed in the next section. \citet{mackenty00} also notice a large number of bright stars outside the main parts of the SF regions in the H$\alpha$ continuum image across the entire galaxy, suggesting that these UV-bright stars are not artifacts. 

We find a strong spatial correlation between the derived $A_{\rm V}$ map and the observed 24~$\mu$m emission. Although the 24~$\mu$m observation has lower angular resolution, it is clear that higher $A_{\rm V}$ values are found in the same regions where the 24~$\mu$m emission is stronger. While the overall dust extinction in NGC 4214 is not severe \citep[$E(B-V) <$ 0.1;][]{leitherer96, mackenty00, ubeda07b, lee09}, the extinction is significant in the main SF complexes and some of the small SF knots. There are some spatial bins with high $A_{\rm V}$ values well outside prominent SF regions, and they are mostly contributed from stars with high $A_{\rm V}$ uncertainty. However, some of those bins actually coincide with weak ($<$ 1--2~MJy~sr$^{-1}$) 24~$\mu$m emission, which is below our minimum contour level, probably indicating clumpy ISM structure in the outer galaxy.

The derived $A_{\rm V}$ map is a useful first-order diagnostic of the ISM structure, giving insights into how ionizing photons may escape. The top right panel in Figure~\ref{avMaps} shows a zoom-in on the $A_{\rm V}$ map around the Central and NW SF complexes (see Figure~\ref{n4214comp} for the region definition). The dust distribution in these regions reflects the prominent shell structures seen in H$\alpha$ (bottom right panel) -- low extinction in the center and high extinction at the boundaries of the SF complexes, indicating that the dust and the ionized gas are physically associated. \citet{maiz-apellaniz98} also found that the dust in the NW complex is decoupled from the central star clusters, and is located at the boundary. This type of dusty shell is produced when strong stellar feedback expels the remaining ISM from the nascent molecular clouds. The $A_{\rm V}$ map, therefore, supports spatial decoupling between evolved (i.e., the first supernovae have erupted) massive star clusters and the surrounding ISM \citep[e.g.,][]{maiz-apellaniz98, mackenty00}. This process also naturally leads to optically-thin holes through which ionizing photons from the young massive stars can escape. Thus, one can expect high escape fractions from the vicinity of these SF complexes.

\subsection{Distribution of Massive Hot Stars}\label{sec:mass_temp}

\begin{figure}
\centering
\includegraphics[scale=0.5, trim=0cm 2cm 0cm 3cm, clip=true]{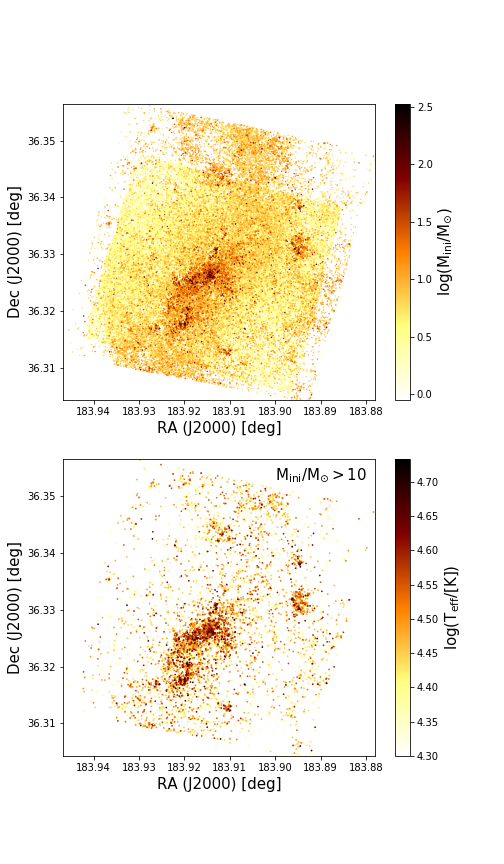}
\caption{Maps of stellar initial mass and temperature. The top panel shows the stellar initial mass distribution. The bottom panel shows the stellar temperature distribution of stars more massive than 10~\msun. The smaller yellow area in the upper panel represents the large number of low-mass RGB stars detected in the WFC3/IR field of view, but that are undetected in the larger WFC3/UVIS observations. 
\label{mass_temp}}
\end{figure}

The top panel of Figure~\ref{mass_temp} presents the map of initial mass for all stars. Most massive stars are confined to the three main SF complexes in the galactic center where H$\alpha$ is also most strongly emitted. This trend is consistent with the enhanced recent SF in the galactic central regions \citep{williams11}. The central concentration of massive stars should be imprinted in the stellar temperature map as well. We select massive stars ($M_{\rm ini} \geq$ 10\msun) and examine the spatial distribution of their effective temperature (bottom panel). As expected, the hottest stars are also clustered mostly in the central SF complexes and a couple of SF knots. 

These highly-clustered hot, massive (and thus young) stars are the primary ionizing sources in NGC 4214. That means the ionizing photon production in NGC 4214 is highly localized. These stars are also spatially correlated with the ionized gas tracer, H$\alpha$ emission (Figure~\ref{ancillary}), and the warm dust tracer, 24~$\mu$m emission (Figure~\ref{avMaps}). This correlation suggests that both the photoionization and gas/dust absorption are localized processes, implying significant local variation in the escape fraction as well. To trace the local variation in $f_{\rm esc}$, we divide NGC 4214 into 13 prominent SF regions based on the H$\alpha$ emission strength and morphology (see Figure~\ref{ancillary}), and measure the local escape fractions for 13 individual SF regions. We will discuss these measurements in detail in Section~\ref{sec:fesc}.  

To get a rough idea of the evolutionary stage of the three primary SF complexes (NW, Central, and SE), we examine the median age of their young stars ($<$ 10~Myr). The NW, Central, and SE complexes show median ages of 5.0, 5.6, and 4.6~Myr, respectively. Thus, one can expect youngest stellar populations in the SE complex and slightly older populations in the Central complex than the NW and SE complexes. This trend is consistent with the results from the stellar cluster analysis \citep{ubeda07b} and composite spectrum analysis \citep{leitherer96}. \citet{sollima14} also recovered the SFH for the Central complex and reported a strong peak of SFR at age of $\sim$8~Myr. 
In addition, the compact H$\alpha$ morphology of the SE complex suggests that it is in an early evolutionary stage. No significant spatial decoupling between the ISM and embedded stars would prevent ionizing photons from escaping the complex. Thus, we expect a higher fraction of ionizing photons escaping from the Central complex than the SE complex. 

\begin{figure} 
 \centering
     \includegraphics[scale=0.5, trim=0.5cm 2cm 1cm 3cm, clip=true]{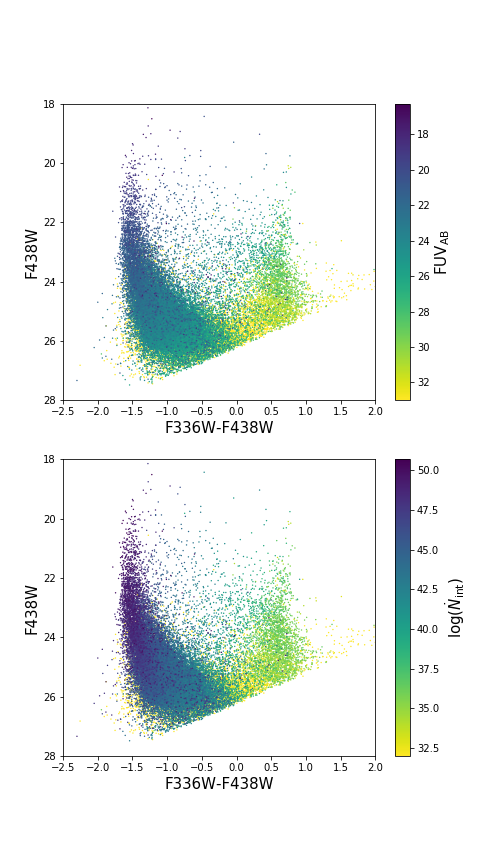} 
       \caption{CMDs color-coded by the BEAST-predicted FUV magnitude (top) and by the intrinsic ionizing photon production rate $\dot{N}_{\rm int}$ (bottom). This figure describes how many FUV and LyC photons are produced by individual stars. As expected, bright young MS stars produce the majority of the FUV and LyC fluxes.
\label{cmd_fuvLyC}}
\end{figure}    

\subsection{Verifying the LyC Flux Production}\label{fuv_lyc}

Figure~\ref{cmd_fuvLyC} shows the CMDs color-coded by the BEAST-predicted unreddened FUV magnitude (top panel) and the intrinsic ionizing photon production rate $\dot{N}_{\rm int}$ (bottom panel). We infer intrinsic and attenuated fluxes at the GALEX FUV and LyC regimes from the SED fits of individual stars by convolving their model SEDs with the GALEX FUV bandpass and a LyC tophat filter, which ranges from 228~\AA~to 912~\AA\,\footnote{The contribution of high-energy photons with $\lambda <$ 228~\AA, corresponding to He$\,\textsc{ii}$ ionization potential of $\sim$54.42~eV (4~Rydberg), to the photoionization of hydrogen is negligible due to their small ionization cross-section.}. As expected, brighter MS stars emit more FUV and LyC fluxes. Even among MS stars, there is significant variation in FUV flux and $\dot{N}_{\rm int}$ from the hottest to the coolest stars. Furthermore, all RGB stars produce at least $\sim$10$^{5}$ times fewer ionizing photons per second than the coolest MS stars in our sample. Thus, it is clear that bright MS stars are responsible for most of the FUV and LyC fluxes in NGC 4214. Although these bright MS stars have high completeness, we further validate whether the stars in our sample produce essentially all of LyC flux in NGC 4214. 

\begin{figure} 
 \centering
      \includegraphics[width=9cm, trim=0cm 0cm 0cm 1cm, clip=True]{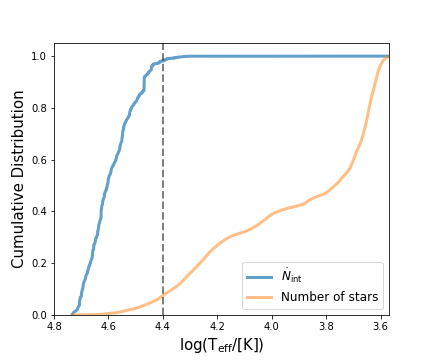}
       \caption{Cumulative distributions of $\dot{N}_{\rm int}$ and the number of contributing stars as a function of log(T$_{\rm eff}$). We draw a dashed line where the cumulative amount of $\dot{N}_{\rm int}$ reaches 99\% (log(T$_{\rm eff}$) $\simeq$ 4.4). Nearly all (97\%) stars hotter than log(T$_{\rm eff}$) $=$ 4.4 are brighter than F438W $=$ 26.25~mag, and completeness at F438W $=$ 26.25~mag reaches $\sim$77\%, suggesting the depth of our data is perfectly adequate for the LyC study although we might miss some highly embedded stars. 
      \label{comp_LyC}}
\end{figure}

\begin{figure*} 
 \centering
     \includegraphics[width=20cm, trim=2.5cm 0cm 0cm 1cm, clip=true]{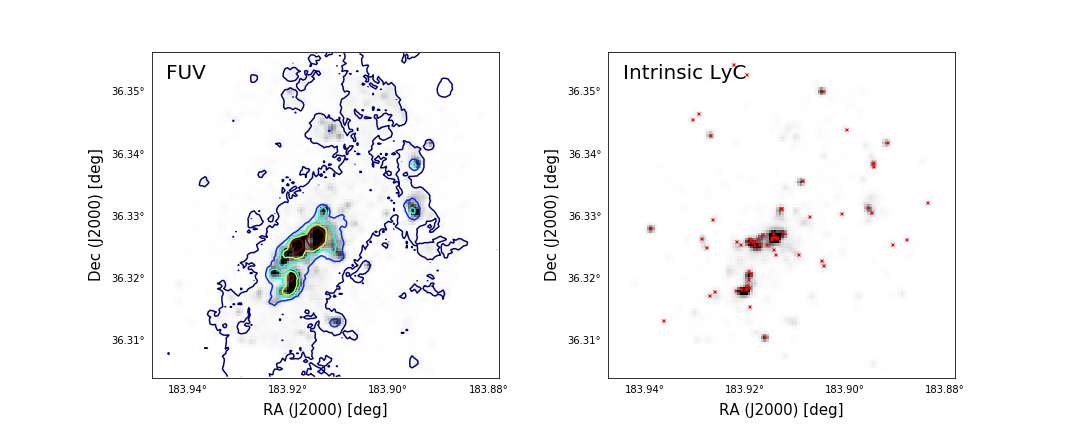}
      \caption{{\it Left:} Comparison of the reconstructed FUV map (gray scale), derived from the BEAST SED fitting of individual stars, with actual GALEX FUV observations (contours) at the GALEX FUV angular resolution. They show an excellent agreement in their local and global morphology. {\it Right:} The predicted map of the intrinsic ionizing photon production rate ($\dot{N}_{\rm int}$) at the GALEX FUV angular resolution. Red crosses denote the 73 stars with $M_{\rm ini} >$ 100~\msun\, and log(T$_{\rm eff}$) $\gtrsim$ 4.4. These stars produce about a quarter of the total $\dot{N}_{\rm int}$ of NGC 4214. Both panels have been smoothed with a large Gaussian filter to match GALEX FUV resolution of 4.2$''$. 
      \label{map_fuv_LyC}}
\end{figure*}

Figure~\ref{comp_LyC} shows the cumulative distribution of $\dot{N}_{\rm int}$ produced by individual stars, as a function of log(T$_{\rm eff}$). Stars with log(T$_{\rm eff}$) $\gtrsim$ 4.4, corresponding to early B-type stars, are responsible for 99\% of $\dot{N}_{\rm int}$. The fraction of these hot stars in our catalog is only $\sim$7\%, and nearly all of them (97\%) are brighter than F438W $=$ 26.25~mag. In each of the eight stellar number density bins (see Section~\ref{sec:noise}), the 50\% completeness is achieved at magnitudes fainter than F438W $=$ 26.25~mag based on extensive ASTs. At F438W $=$ 26.25~mag, the completeness reaches for stars of all temperatures at all locations is $\sim$77\%.  

There is a reasonable question as to whether our data could miss some of the most deeply embedded hot stars, leading to underestimations of the true $\dot{N}_{\rm int}$. The noise model indicates that our data can probe stars with extinction up to $A_{\rm V} \simeq$ 5~mag, but that we rarely recover stars with $A_{\rm V} >$ 5~mag at any temperature. Therefore, it is likely that highly embedded young stars are not included in our sample. 

That said, even if we were to miss some heavily attenuated stars, the number of escaping LyC photons would be unaffected because none of the ionizing photons from those non-detected stars would contribute to the emerging LyC flux. The escape fraction measurement would not be significantly affected either because those stars compose only a minor fraction of the young, hot stars in NGC 4214 based on the fact that NGC 4214 has low internal extinction \citep[e.g.,][]{leitherer96}. Indeed, the $A_{\rm V}$ distribution of hot stars (log(T$_{\rm eff}$) $\gtrsim$ 4.4) shows that about 98\% (95\%) of them have $A_{\rm V} <$ 3 (2) with a mean and median of 0.56 and 0.33, respectively. 

To further investigate our recovery of the {\it spatial} distribution of non-ionizing UV flux, we compare the map of reddened FUV fluxes in predicted by the BEAST (gray scale) with the actual GALEX FUV observation (contours) in the left panel of Figure~\ref{map_fuv_LyC}. For direct comparison with GALEX FUV, we degrade the high resolution FUV map constructed from the BEAST-predicted FUV fluxes of individual stars by convolving with a Gaussian kernel with the GALEX FUV image's FWHM of 4.2$''$ to match their angular resolution. 

The FUV GALEX observation and the BEAST prediction agree very well morphologically. The predicted FUV map recovers both the global shape and the small structures for individual SF complexes seen in the GALEX image. The excellent morphological agreement between the observation and our prediction suggests reliable SED modeling for stars that are indeed sources of FUV flux. 

We note, however, that the HST data are not deep enough to be 100\% complete for FUV sources. From the predicted FUV map, we compute the total magnitude by considering only the pixels 3$\times$ above the GALEX FUV detection threshold for making a fair quantitative comparison with the GALEX observation. The estimated total FUV magnitude is $\sim$12.5~mag, which is $\sim$1.57 times fainter than the GALEX observed flux within the same areal coverage ($\sim$12.01~mag\footnote{This is 0.54~mag fainter than the measurement from \citet{lee11} for NGC 4214, simply because we consider a smaller area than their photometric area.}). Most of missing FUV flux compared to the GALEX observation comes from the extended disk surrounding the central SF complexes, indicating that our data are too shallow to detect the older populations (late B-type and A-type), which also produce FUV flux, but do not produce a significant amount of ionizing flux (see Figure~\ref{comp_LyC}). This difference in FUV is expected from the rather shallow depth of the image and our 4-band requirement for more reliable SED fitting. Scattered light might also partially explain the discrepancy. The integrated light of a dusty galaxy inevitably contains a certain fraction of scattered light. A fraction of scattered to stellar flux at the FUV wavelength could range $\sim$10-50\% for a galaxy with the clumpy ISM \citep{witt00}.

Nevertheless, the excellent morphological agreement between our FUV prediction and the GALEX FUV observation further validates our method to infer LyC flux because it implies that our reconstruction is likely to hold at shorter wavelengths where the completeness is higher than the FUV regime. We emphasize that our completeness in the LyC regime is highly sufficient for this study, especially when $A_{\rm V} <$ 2~mag. 

The right panel of Figure~\ref{map_fuv_LyC} presents the $\dot{N}_{\rm int}$ map with the same spatial resolution as the GALEX FUV map for easy comparison. Morphologically, the $\dot{N}_{\rm int}$ map correlates with the FUV map, confirming that most FUV bright stars also produce ionizing photons. It is clearly seen that production of ionizing photons in the galaxy is highly localized, as expected from the maps of stellar mass and temperature (Figure~\ref{mass_temp}). We overplot the 73 stars with $M_{\rm ini} >$ 100~\msun\, among hot stars with log(T$_{\rm eff}$) $\gtrsim$ 4.4. These massive and hot stars produce $\sim$25\% of the total $\dot{N}_{\rm int}$ of NGC 4214.

\begin{figure*}
\centering
\includegraphics[width=18cm, trim=5cm 0cm 3cm 0cm, clip=true]{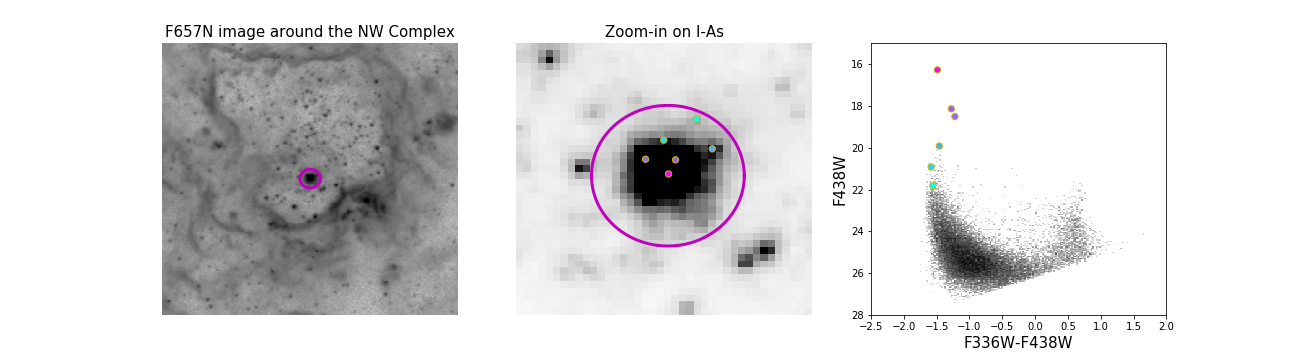}
\caption{{\it Left:} H$\alpha$ (F657N) image cutout around the NW complex. The purple circle indicates the unresolved star cluster I-As with a radius of 0.41$''$. {\it Middle:} Zoom-in on I-As. The colored points are resolved stars detected in at least 4 bands. They are color-coded by brightness in F438W, and can be identified in the right panel. The star in the center of I-As is brighter than the other five stars in all 6 bands, and could be a Wolf-Rayet star \citep[e.g.,][]{sargent91,kobulnicky96,leitherer96,maiz-apellaniz98}. \citet{ubeda07b} predict one Wolf-Rayet star in I-As for the model with metallicity of 0.004 by analyzing the cluster's integrated fluxes. However, severe crowding in the cluster center prevents us from completely ruling out a possibility that this brightest object is merely a blend of multiple bright stars although this object satisfies the photometric quality cuts in all 6 bands. {\it Right:} Position of the resolved six stars on the CMD. The color scheme is identical to the one used in the middle panel. 
\label{I-As}}
\end{figure*}

\section{The Escape Fraction Measurement} \label{sec:fesc}
In a dusty HII region, gas and dust compete with each other to absorb LyC photons. We thus need the ionizing photon production rate ($\dot{N}_{\rm int}$) and the ionizing photon consumption rates both by neutral hydrogen gas ($\dot{N}_{\rm gas}$) and dust ($\dot{N}_{\rm dust}$) to measure $f_{\rm esc}$. Once we have all these measurements for a region of interest, we can compute the ionizing photon escape fraction as follows:
\begin{equation} \label{eq_fesc}
\begin{split}
f_{\rm esc} & = \frac{
     \dot{N}_{\rm int} - \dot{N}_{\rm gas} - \dot{N}_{\rm dust}}
  {\dot{N}_{\rm int}} \\
  & = 1 - \frac{\dot{N}_{\rm gas}}{\dot{N}_{\rm int}} - \frac{\dot{N}_{\rm dust}}{\dot{N}_{\rm int}}.
\end{split}
\end{equation}
The first two terms, $\dot{N}_{\rm int}$ and $\dot{N}_{\rm gas}$, are independent of viewing angle and thus readily calculated. The first term can be measured by combining ionizing photons produced by all stars in a region of interest (either individual HII regions or a whole galaxy), and the second term can be measured from the isotropically emitted extinction-corrected H$\alpha$ recombination emission-line luminosity of the region. 

On the other hand, measuring $\dot{N}_{\rm dust}$ is challenging because both dust properties and the relative star/dust geometry vary along sightlines, and we only know the dust absorption along our line of sight from the BEAST fitting. As a result, the total $\dot{N}_{\rm dust}$ that is estimated solely based on the dust properties measured from the BEAST fitting would be completely different from those measured by observers in other directions. We instead employ a dust ``covering factor'' ($f_{\rm cov}$) -- the fraction of HII region's surface covered by the optically-thick ISM to LyC photons to account the amount of LyC photons absorbed by dust before escaping a SF region (see Section~\ref{sec:fcov} for the detailed discussion). 

Fortunately, a majority of LyC absorption in metal-poor dwarf galaxies is likely attributed to neutral hydrogen gas rather than the dust. First, the fraction of LyC photons contributing to hydrogen ionization in the HII region is known to anti-correlate with the dust-to-gas ratio or metallicity \citep[e.g.,][]{mathis71, petrosian72, inoue01}. According to these studies, the fraction of LyC photons absorbed by neutral hydrogen in NGC 4214 is expected to be high due to its low metallicity and corresponding low dust content. In fact, NGC 4214 has a very high gas-to-dust ratio of 350--470 \citep{hermelo13}. Second, the absorption cross-section of dust per hydrogen nucleus is smaller than the ionization cross-section of hydrogen by a factor of $\sim$10$^2$ at 228~\AA\,and $\sim$10$^4$ at 912~\AA, respectively, indicating a much larger likelihood of LyC being absorbed by neutral hydrogen gas over dust \citep[see Fig.2 of][]{glatzle19}. 

We also do not expect a major impact due to helium absorption. Neutral helium is not the primary gas component competing with the dust grains in the wavelength regime 228~\AA\, $ < \lambda <$ 912~\AA. This is because even if a portion of high energy ($\lambda <$ 504~\AA) photons is used to ionize neutral helium rather than hydrogen, most recombinations from singly ionized helium produce hydrogen ionizing photons \citep{mathis71}. Furthermore, the helium abundance in NGC 4214 is low relative to hydrogen \citep[0.0842 $\pm$ 0.0019;][]{kobulnicky96}.

In total, it is reasonable to expect that emitted LyC photons in 228~\AA\, $ < \lambda <$ 912~\AA\,are predominantly absorbed by neural hydrogen gas. Any residual LyC photons are absorbed by dust before leaving the SF region or the galaxy. This assumption makes the effect of complex ISM geometry simple in the $f_{\rm esc}$ measurement.

\begin{figure*} 
 \centering
     \includegraphics[width=18cm, trim=3cm 0.5cm 3cm 1.5cm, clip=true]{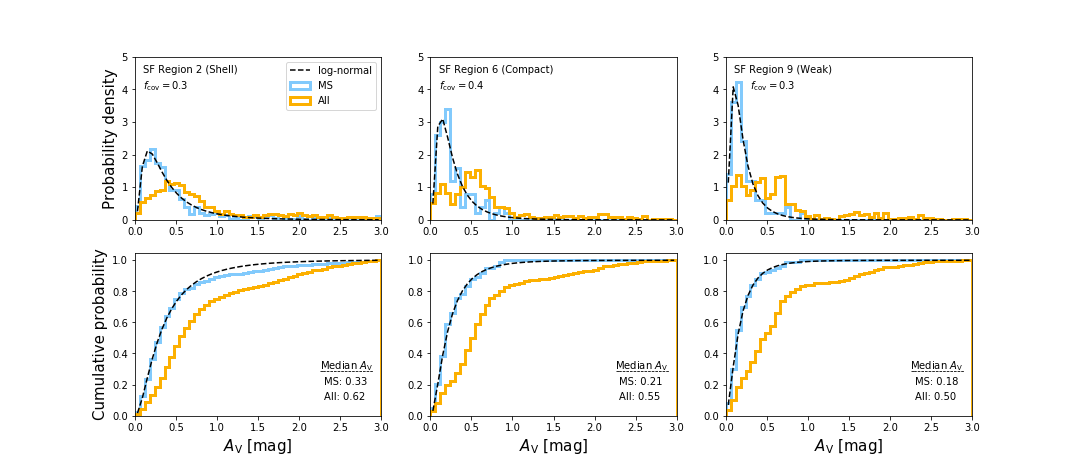}
      \caption{Top: The measured $A_{\rm V}$ distributions of three SF regions (SF regions 2, 6 and 9) representing different HII region's morphologies. The $A_{\rm V}$ distribution of young MS stars is shown in light blue, while that of all stars in each SF region shown in orange. The black dashed lines show the best-fitting lognormal function to each MS distribution. We note that the median $A_{\rm V}$ in each region is low, and that there is no noticeable indication of a significant fraction of highly embedded ($A_{\rm V} >$ 3) population of MS stars. The median $A_{\rm V}$ values for young MS and all stars in each SF region are given in the bottom panels.  
      \label{Av_distribution}}
\end{figure*}

\subsection{Intrinsic Ionizing Photon Production Rate}\label{calc_nint}
We take the following steps to estimate the total $\dot{N}_{\rm int}$ and its uncertainty attributed from SED modeling for each SF region. The BEAST provides 1D pPDFs for $\dot{N}_{\rm int}$ for individual stars. In a given SF region, we randomly sample $\dot{N}_{\rm int}$ 10,000 times for individual stars inside that SF region based on each star's 1D pPDFs, resulting in 10,000 measurements of $\dot{N}_{\rm int}$ for the given SF region. From the distribution of these 10,000 measurements, we compute 16th, 50th, and 84th percentile values to obtain the median value of $\dot{N}_{\rm int}$ and its asymmetric 1$\sigma$ uncertainties. These asymmetric uncertainties are also reported in Table~\ref{tb_fesc}. 

\subsubsection{Contribution of Unresolved Objects to $\dot{N}_{\rm int}$}
NGC 4214, at its distance, inevitably has unresolved star clusters \citep[e.g.,][]{mackenty00}. Thus, the contribution of unresolved star clusters to $\dot{N}_{\rm int}$ needs to be considered as well. Because the BEAST is designed to perform on resolved stars and our analysis is limited to objects in the good star catalog (Section~\ref{sec:data}), the BEAST results likely underestimate the true $\dot{N}_{\rm int}$ by missing ionizing photons contributed by unresolved objects. Note, however, that only clusters with ages less than 10~Myr will contribute to the production of ionizing photons.  

\citet{ubeda07b} analyze star clusters in NGC 4214 using \textit{HST} WFPC2 and STIS photometry -- four of them are unresolved star clusters \citep[I-As, I-Es, IIIs, and IVs; see their positions and size in][]{ubeda07a}. Among these unresolved star clusters, I-As (the super star cluster in the SF region 1) is the only young cluster that can produce significant amount of LyC photons. I-Es might produce some LyC photons if it is indeed a young cluster; there are two age solutions (7.1~Myr vs. 189~Myr) with a higher probability for the older-age solution \citep{ubeda07b}. However, even if I-Es is a young cluster, its contribution to $\dot{N}_{\rm int}$ is expected to be negligible compared to I-As based on its age and the predicted number of O/B stars \citep{ubeda07b}.

Figure~\ref{I-As} presents a compact ionizing star cluster I-As \citep[radius of 0.41$''$;][]{ubeda07a} on the H$\alpha$ image and some of its resolved stars on the CMD. With the WFC3 imaging data and our simultaneous six-filter photometry, we are able to resolve ten individual stars, which pass our photometric quality cuts, within I-As. Thus, I-As is not a completely unresolved star cluster, but a partially resolved star cluster in our photometry. Six out of ten stars meet the 4-band detection requirement and thus have the SED fitting results. These six stars are hot and young based on their observed SEDs and the BEAST SED fits; two O-type stars, three blue supergiants, and one likely Wolf-Rayet star from fainter to brighter in F438W. In fact, the Wolf-Rayet features have been observed in the spectrum of I-As and the spectral features have indicated the presence of both nitrogen(WN)- and carbon (WC)-sequence Wolf-Rayet subtypes \citep{sargent91, kobulnicky96, leitherer96, maiz-apellaniz98}. 

In the middle panel of Figure~\ref{I-As}, the pink object in the center of I-As is the brightest object in our catalog in F225W, F336W, F438W, F814W, and F110W, and its observed SED is about 2$\times$ brighter than any of our model SEDs in the two UV bands, resulting in unsuccessful SED modeling for this particular star. Specifically, the BEAST inappropriately models this object as a massive red supergiant, making it the least important LyC contributor among the six resolved stars. If this object is a Wolf-Rayet star with a high mass-loss rate, its observed SED can not be optimally modeled by our stellar models because the non-LTE TLusty atmosphere models used in this study are plane-parallel and hydrostatic. Alternatively, crowding at the distance of NGC 4214 prevents us from excluding the possibility that this object is a composite of multiple bright sources even though the object passes our conservative photometric quality cuts in all 6 bands. In either cases, our $\dot{N}_{\rm int}$ prediction for this object unambiguously underestimates the true value. With the `incorrect' minimum contribution of the brightest pink object in mind, the combined $\dot{N}_{\rm int}$ of resolved six stars is $\sim$5$\times$10$^{50}$~s$^{-1}$.

The earlier composite spectroscopic studies of I-As have predicted a few to a couple of Wolf-Rayet stars in I-As when assumed the same distance to NGC 4214 of $\sim$3~Mpc \citep[e.g.,][]{leitherer96,maiz-apellaniz98,ubeda07b}. Based on spectral analysis of I-As with a 1$''$ aperture (very similar to the size of I-As used in this study), \citet{leitherer96} infer $\dot{N}_{\rm int} =$ 1.9$\times$10$^{51}$~s$^{-1}$ and 4.1$\times$10$^{51}$~s$^{-1}$ for the burst and continuous star formation models, respectively, assuming the distance to NGC 4214 of 4.1 Mpc. These can be translated to $\sim$1.04$\times$10$^{51}$~s$^{-1}$ and 2.25$\times$10$^{51}$~s$^{-1}$ for the burst and continuous star formation models, respectively, at the distance of 3.04~Mpc used in this study. 

To consider maximum contribution of the unresolved (partially resolved in our study) star cluster I-As to $\dot{N}_{\rm int}$ of the SF region 1, we add the difference in $\dot{N}_{\rm int}$ between the continuous SF model case and our six resolved star-based case (i.e, $\sim$1.75$\times$10$^{51}$~s$^{-1}$) to the total $\dot{N}_{\rm int}$ for the SF region 1. This addition allows us to account for $\dot{N}_{\rm int}$ contribution from `unknown' numbers of unresolved ionizing stars in I-As in addition to the six resolved stars. That is, $\dot{N}_{\rm int}$ reported in Table~\ref{tb_fesc} for the SF region 1 is the sum of contribution from I-As as the unresolved star cluster and contribution from individual stars outside I-As. 

The intrinsic ionizing photon production rate of  2.25$\times$10$^{51}$~s$^{-1}$ makes I-As alone contribute to roughly 21\% of the total $\dot{N}_{\rm int}$ of the SF region 1. We recall that I-As is the very compact ionizing star cluster within the NW SF complex in the SF region 1. A comparable ionizing star cluster R136 within the 30 Doradus SF complex in the Large Magellanic Cloud is well studied due to its proximity and also known to harbor four very massive hydrogen-rich WN subtypes (160--320~\msun). These four WN stars provide 46\% of the total ionizing photons produced in R136, and the star cluster R136 is responsible for $\sim$40\% ($\sim$4$\times$10$^{51}$~s$^{-1}$) of the total ionizing photons produced the 30 Doradus SF complex \citep{crowther10}. Despite the comparable sizes of I-As and R136, R136 produces at least $1.8\times$ more ionizing photons than I-As primarily due to its very young age of 1.5~Myr \citep{crowther16}. Both the I-As and R136 cases confirm highly localized distribution of ionizing sources.

\subsubsection{Contribution of Highly Extinguished Stars to $\dot{N}_{\rm int}$}
As mentioned Section~\ref{fuv_lyc}, the nature of our relatively shallow data and the 4-band detection requirement prevents us from a complete census of ionizing stars. Although a fraction of undetected, highly embedded stars is expected to be significantly low in NGC 4214 due to its low dust content, the ASTs indicates that we likely miss some highly extinguished young MS stars ($A_{\rm V} >$ 3). Inferring the exact amount of missing $\dot{N}_{\rm int}$ due to high dust columns along our sightlines is non-trivial. To correctly estimate the total $\dot{N}_{\rm int}$ for a highly embedded SF region, one might, at least, need a precise recent SFH with high time resolution and the effect of stochastic sampling of the IMF on the $\dot{N}_{\rm int}$ needs to be taken into account as well, which is beyond the scope of this paper.

Instead of completely ignoring missing ionizing photons from highly extinguished stars, we estimate the minimum number of missing ionizing photons in the SF regions where the BEAST predicted $\dot{N}_{\rm int}$ is smaller than $\dot{N}_{\rm gas}$, leading to a negative local $f_{\rm esc}$, by requiring $f_{\rm esc}$ = 0 (i.e., all LyC photons are consumed to ionize the neutral hydrogen gas). Out of 13 SF regions, there is only one SF region showing a negative $f_{\rm esc}$ -- the SF region 3b. We measure $\dot{N}_{\rm int} =$ 2.66$\times10^{50}$~s$^{-1}$ for the SF region 3b from the BEAST fits, while its $\dot{N}_{\rm gas}$ is measured to be 4.95$\times10^{50}$~s$^{-1}$. It is not surprising that we underestimate the true $\dot{N}_{\rm int}$ based on its highest mean and standard deviation values of the lognormal best-fit function to the $A_{\rm V}$ distribution (see Section~\ref{sec:fcov}). To make $f_{\rm esc}$ = 0, $\dot{N}_{\rm int}$ needs to balance with $\dot{N}_{\rm gas}$ and we thus replace the BEAST predicted $\dot{N}_{\rm int}$ with $\dot{N}_{\rm gas}$ of 4.95$\times10^{50}$~s$^{-1}$.

\begin{deluxetable*}{lccccccccccccc}
\tablecaption{$A_{\rm V}$ Distribution of Young MS Stars for Each SF Region \label{tb_lognorm}} 
\tablewidth{0pt}
\tablehead{
\colhead{SF Region} & \colhead{1} & \colhead{2} & \colhead{3a} & \colhead{3b} & \colhead{3c} & \colhead{3d} & \colhead{3e} & \colhead{4} & \colhead{5} & \colhead{6} & \colhead{7} & \colhead{8} & \colhead{9} \\
\hline
Other Name\tablenotemark{a}  & I-A/I-Cn/I-D & I-B & II-C & II-B & II-Dn & II-En & II-A & IXn & VIn & Vn & -- & VIIn/XIIIn & VIIIn
}
\startdata
$\mu$~($A_{\rm V}$) & 0.31 & 0.30 & 0.22 & 0.34 & 0.25 & 0.23 & 0.32 & 0.24 & 0.20 & 0.21 & 0.16 & 0.22 & 0.19 \\
$\sigma~(A_{\rm V})$ & 0.77 & 0.75 & 0.85 & 1.19 & 0.81 & 0.82 & 0.85 & 0.51 & 0.65 & 0.58 & 1.23 & 0.66 & 0.50
\enddata
\tablecomments{Fitting a lognormal function to the $A_{\rm V}$ distribution of young MS stars in each SF region: $\exp{(-(\ln{x}-\mu)^{2}/2b^2)}/(x\sigma\sqrt{2\pi}$).}
\tablenotetext{a}{HII knots identified by \citet{mackenty00} that roughly correspond to our SF regions.}
\end{deluxetable*}

\subsection{Gas Absorption Traced by H$\alpha$}\label{calc_ngas}
We account for the neutral gas absorption using the extinction-corrected H$\alpha$ imaging. The absorption of LyC photons by neutral hydrogen gas is followed by the subsequent emission of hydrogen recombination lines arising from ionized gas, and this emission spreads out isotropically. Therefore, the ionizing photon consumption rate by photoionization ($\dot{N}_{\rm gas}$) can be measured through the H$\alpha$ luminosity. The conversion factor between H$\alpha$ luminosity and the corresponding ionizing photon production rate is well established for standard radiative transfer assumptions \citep{kennicutt98}. Under the assumption of Case B recombination (i.e., the on-the-spot approximation) for $ T_{\rm e} = 10^{4}$~K and $n_{\rm e} = 10^{2}$~cm$^{-3}$ \citep{hummer87,storey95}, 
\begin{equation} \label{haToNion}
\dot{N}_{\rm gas} = 7.315\times10^{11}L(\rm H\alpha).
\end{equation} 

We derive an extinction-corrected H$\alpha$ map using HST narrow-band images of H$\alpha$ (F657N) and H$\beta$ (F487N) lines. We first remove stellar continuum from the H$\alpha$ narrow-band image. The stellar continuum is computed by interpolating the F547M and F814W images at the effective wavelength of the F657N filter, and then we subtract the scaled H$\alpha$ stellar continuum image from the F657N narrow-band image. For continuum subtraction in the H$\beta$ narrow-band image, we use the scaled F438W image. A scaling factor for each continuum image was determined by minimizing absolute residuals in regions where nebular emission is negligible. 

We also account for the [NII] contribution to the F657N using the [NII]/H$\alpha$ ratio inferred based on the H$\alpha$ intensity in each pixel. \citet{maiz-apellaniz98} analyze the spatially resolved optical spectrum of NGC 4214 and map various emission lines and their ratios, including H$\alpha$ and [NII]/H$\alpha$. They find a tight correlation between these two values in their pixel by pixel comparison both in the NW and SE Complexes (see their Fig. 11). Unfortunately, in the absence of a data table or vector graphic information, we have resorted to
using a by-eye fit to the data points. Since the NW and SE Complexes show a very similar trend, we obtain a single fit that describes the both Complexes reasonably well: 
\begin{equation} \label{eq_nii}
[NII]/H\alpha = -0.057\times\log(I(H\alpha)) - 0.731,
\end{equation} 
where $I$(H$\alpha$) is the H$\alpha$ intensity in units of erg~s$^{-1}$~cm$^{-2}$~arcsec$^{-2}$.

With these continuum-subtracted H$\alpha$ and H$\beta$ emission images, we derive a dust reddening from the Balmer decrement in each pixel, and apply pixel-by-pixel dust correction to the H$\alpha$ emission-line map. Dust correction based on the Balmer decrement is known to underestimate the true H$\alpha$ flux because the Balmer decrement-based optical depth reflects the true optical depth at H$\alpha$ only for the foreground uniform dust screen gas/dust geometry or low attenuation circumstances \citep[e.g.,][]{caplan86}. In reality, an HII region has a complex relative geometry of the gas and dust in 3D space (see Figure~\ref{avMaps}), and the observed total emission from the mixture of the gas and dust is luminosity weighted, biasing the extinction measurement to lower values. In addition, the scattered-in photons can lower the Balmer ratio when dust clouds exist near the emitting gas, which is likely the case for the gas/dust mixture geometry \citep{maiz-apellaniz98}. Therefore, we make an additional correction to the measured H$\alpha$ flux to account for the complex gas/dust geometry in each SF region as follows.

The true optical depth at H$\alpha$ can be estimated using the thermal free-free radio continuum, which is produced by the same electrons that produce H$\alpha$ emission, but is unaffected by dust. \citet{maizapellaniz04} show that the optical depth from the thermal radio continuum is always greater than that from the Balmer decrement in NGC 604, a giant HII region in M33. The discrepancy between the two optical depths becomes zero only for the uniform dust screen case, and increases for all more complex gas/dust geometry cases. They also compare the optical depths from the Balmer decrement method, but measured at different spatial resolutions (integrated light within an aperture of a SF region vs. correcting for extinction pixel by pixel). The pixel by pixel correction returns a significantly larger optical depth than the integrated flux case by 20-130\%, and thus closer to the true optical depth values. Finally, they find an underestimation in H$\alpha$ luminosity by $\sim$11\% for the pixel by pixel correction case and by $\sim$27\% for the integrated light correction case in NGC 604. This suggest that our measured H$\alpha$ flux corrected for dust pixel by pixel would not be significantly different from the true values. 

In NGC 4214, \citet{mackenty00} conduct a similar analysis, but on the limited number SF knots due to lower angular resolution of radio data. They also find larger optical depths from radio data than those from the Balmer decrement (taking the ratio of integrated H$\alpha$ and H$\beta$ fluxes). The lack of information on the true optical depths for individual SF regions, significant uncertainties due to contributions from nonthermal components, and the difference in the angular resolution for Balmer decrement measurements (integrated light vs. pixel by pixel) hamper us from utilizing the optical depth values based on the radio data from their study. 

On the other hand, \citet{mackenty00} provide the extinction-corrected H$\alpha$ fluxes for individual SF knots in NGC 4214 using the integrated flux-based Balmer decrement assuming the dust screen and gas/dust mixture\footnote{No details about the methodology are available. They refer an `in preparation' paper that seems not published.}. With variation in individual SF regions, globally, the dust screen correction underestimates the H$\alpha$ luminosity by $\sim$24\% compared to the gas/dust mixture correction. If their gas/dust mixture correction is close enough to the case reflecting the true optical depth, this is consistent with the underestimate of $\sim$27\% for the integrated light correction case seen in NGC 604. Thus, we adopt the underestimation of 11\% in H$\alpha$ flux for the pixel by pixel correction found in \citet{maizapellaniz04} and apply to the entire galaxy to correct for the gas/dust geometry effect on the inferred H$\alpha$ luminosity. Using the distance of 3.04~Mpc, we obtain the H$\alpha$ luminosity and convert this to the ionizing photon absorption rate by the surrounding neutral hydrogen ($\dot{N}_{\rm gas}$) using Equation~\ref{haToNion}.

The converted total number of ionizing photons per second is $\dot{N}_{\rm gas} \simeq~1.58\times10^{52}$~s$^{-1}$ with $\sim$20\% uncertainty. The uncertainty in $\dot{N}_{\rm gas}$ includes the uncertainties in removing the stellar continuum contribution to H$\alpha$ and H$\beta$ emission, and correcting for the [NII] contribution to H$\alpha$ flux. Our $\dot{N}_{\rm gas}$ is in good agreement with the thermal radio continuum-based measurements. For example, the $\dot{N}_{\rm gas}$ prediction based on the flux density at 6~cm is $\sim$1.11--1.83$\times$10$^{52}$~s$^{-1}$ depending on the fraction of thermal component of 0.4--0.66 at 6~cm \citep[e.g.,][]{kepley11,hindson18}. We make the conversion between the thermal radio continuum and the ionizing photon production rate following \citet{condon92}.

The measurements of $\dot{N}_{\rm gas}$ for the 13 SF regions are presented in Table~\ref{tb_fesc}. We use these values to calculate the local escape fraction of each SF region in Section~\ref{sec:localEsc}. The combined $\dot{N}_{\rm gas}$, or equivalently H$\alpha$ luminosity, of the 13 SF regions contributes significantly ($\sim$78\%) to the total integrated $\dot{N}_{\rm gas}$ over the entire galaxy. The remaining 22\% of H$\alpha$ emission outside these 13 main SF regions can be considered as the diffuse, warm ionized gas. Indeed, \citet{oey07} measured the mean diffuse H$\alpha$ fraction of 36$\pm$18\% for starburst dwarf galaxies, which is a much lower fraction than those for other types of galaxies (e.g., $\sim$60\% for spirals). They suggested that a significant fraction (up to 25\%) of ionizing photons could escape into the IGM rather than being consumed within an outer halo of a starburst galaxy. We note that we use the total integrated $\dot{N}_{\rm gas}$ when calculating the global $f_{\rm esc}$ in Section~\ref{sec:globalEsc} to account for the LyC absorption by diffuse H$\alpha$ before escaping into IGM.

\subsection{Dust Covering Factor} \label{sec:fcov}
When measuring the local $f_{\rm esc}$, we assume each SF region is surrounded by a porous shell of dust (as in Figure~\ref{cartoonEscape}), where a fraction of the shell is covered by the dust that is optically thick to LyC photons. In this model, the effective ISM geometry represents a dust-free HII cavity surrounded by a porous photodissociation region. As we discussed in the previous section, the majority of LyC photons is absorbed by neutral hydrogen first, and any remaining unabsorbed LyC photons are then consumed by dust if they happen to be escaping in a direction covered by dust. We therefore set $\dot{N}_{\rm dust}$ = $f_{\rm cov}\times(\dot{N}_{\rm int} - \dot{N}_{\rm gas})$.

For NGC 4214, \citet{hermelo13} estimated the dust covering factors for two large SF complexes by carefully modeling their dust emission. While their SE region covers the same SF complex as our SE region, their NW region encompasses our combined Central and NW complexes. For a metallicity of 0.2~\zsun, their best-fit models suggest $f_{\rm cov}$ of 0.3 for the Central and NW SF complexes, and they obtained $f_{\rm cov}$ of 0.6 for the SE SF complex. Uncertainties for these best-fit $f_{\rm cov}$ values are not presented in \citet{hermelo13} explicitly. Thus, we use their initial guess ranges of $f_{\rm cov}$ for each complex to assign the uncertainties in $f_{\rm cov}$, which is listed in Table~\ref{tb_fesc}. We will use these uncertainties to compute the lower and upper limits of escaping LyC photons from each SF region when measuring the global escape fraction in Section~\ref{sec:globalEsc}. 

The measurements in \citet{hermelo13} were made based on low angular resolution IR observations (Spitzer, Herschel, Planck, and IRAM), and thus $f_{\rm cov}$ is not available for our individual SF regions. Here, we apply $f_{\rm cov}$ of 0.3 for our NW (region 1) and Central (region2) SF complexes, and the one SF region (region 9) right next to the Central complex as well. For five SF regions in the SE complex (regions 3a--3e), we adopt $f_{\rm cov}$ of 0.6. For the rest of the SF regions without reported $f_{\rm cov}$, we apply 0.4, which is an intermediate value of their initial estimation of $f_{\rm cov}$ ranging from 0.2 to 0.6. We list the adopted $f_{\rm cov}$ values for each SF region in Table~\ref{tb_fesc} along with their measured local $f_{\rm esc}$, $\dot{N}_{\rm gas}$, and $\dot{N}_{\rm int}$. The effect of the variation in the adopted $f_{\rm cov}$ values on the final global $f_{\rm esc}$ will be discussed in the following section.

In NGC 4214, dust emission is also detected in the outside the central SF regions, and \citet{hermelo13} analyzed this dust emission from the diffuse dust component as well by considering both the stellar emission arising from outside the main SF complexes and escaping stellar flux from the SF regions. Their best-fit model suggested that 40--65\% of the emergent flux from the SF regions escapes into the IGM without interacting with the diffuse ISM. We adopt this as 1 -- $f_{\rm cov}$ of the diffuse ISM and apply it when measuring the global $f_{\rm esc}$ in Section~\ref{sec:globalEsc}.

Although the BEAST derived $A_{\rm V}$ values only provide limited information to our sightlines, it may capture the highly structured nature of the ISM. The top (bottom) panels in Figure~\ref{Av_distribution} show the measured (cumulative) $A_{\rm V}$ distributions of young MS and all stars in the SF regions 2, 6 , and 9, respectively. Regardless of the SF region morphology (shell, compact, and weak), the $A_{\rm V}$ distributions of young MS stars are well described by a lognormal distribution that is expected in turbulent-driven ISM structure \citep[e.g.,][]{vazquez-semadeni94,ostriker01,dalcanton15}, while the $A_{\rm V}$ distributions of all stars are broader with peaks at higher extinction. The median $A_{\rm V}$ values for young MS and all stars are presented in the bottom panels. About 2$\times$ larger median $A_{\rm V}$ of all stars indicate that older stars (mostly RGB) likely probe both front and back of the dusty shells, making them to experience longer path length though the ISM, whereas young MS stars mostly residing inside a SF region experience only front of the dusty shells. The mean and standard deviation of the best-fit lognormal function to the $A_{\rm V}$ distribution for each SF region is listed in Table~\ref{tb_lognorm}.

When assuming a physical model where the stars are surrounded by an isotropic dusty ISM shell in all directions, we can use the BEAST measured dust properties towards individual stars to derive an alternative $f_{\rm cov}$. Although the BEAST derived $A_{\rm V}$ values only provide limited information to our sightlines, it may capture the highly structured nature of the ISM. To determine $f_{\rm cov}$ for individual SF regions based on our SED fitting results, we first sample 10,000 sets of $\theta_{\mathrm{dust}} = \{A_V, R_V, f_\mathcal{A} \}$ from the measured dust properties in each SF region. Second, we randomly assign $\theta_{\mathrm{dust}}$ to individual stars in a given SF region, for which we know their intrinsic stellar properties from the SED fitting. We then finally compute the fraction of LyC photons absorbed by the internal dust and take that as $f_{\rm cov}$. This results in 10,000 measurements of $f_{\rm cov}$ for each SF region, from which we measure average $f_{\rm cov}$ for a given SF region. The estimated $f_{\rm cov}$ values for all 13 SF regions solely based on our own SED fitting results range from 0.75 to 0.84 with up to 17\% uncertainty, giving the average of 0.8. This high $f_{\rm cov}$ along our sightline might explain the zero escape fraction reported in the literature for NGC 4214, which used spectral absorption lines also probing the gas and dust only along the line of sight \citep{heckman01,grimes09}.

\begin{figure*}
\centering
\includegraphics[scale=0.59, trim=0cm -0.1cm 1.5cm 1cm, clip=true]{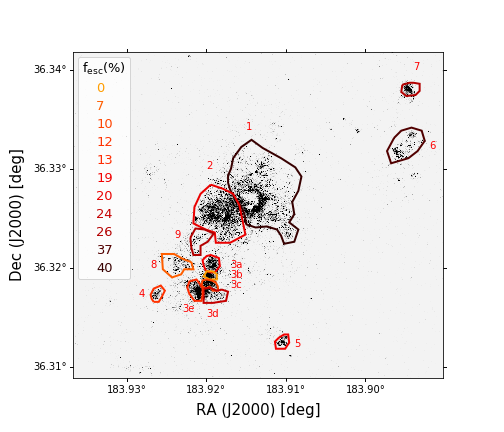}
\includegraphics[scale=0.38, trim=2.5cm -0.5cm 3.5cm 1cm, clip=true]{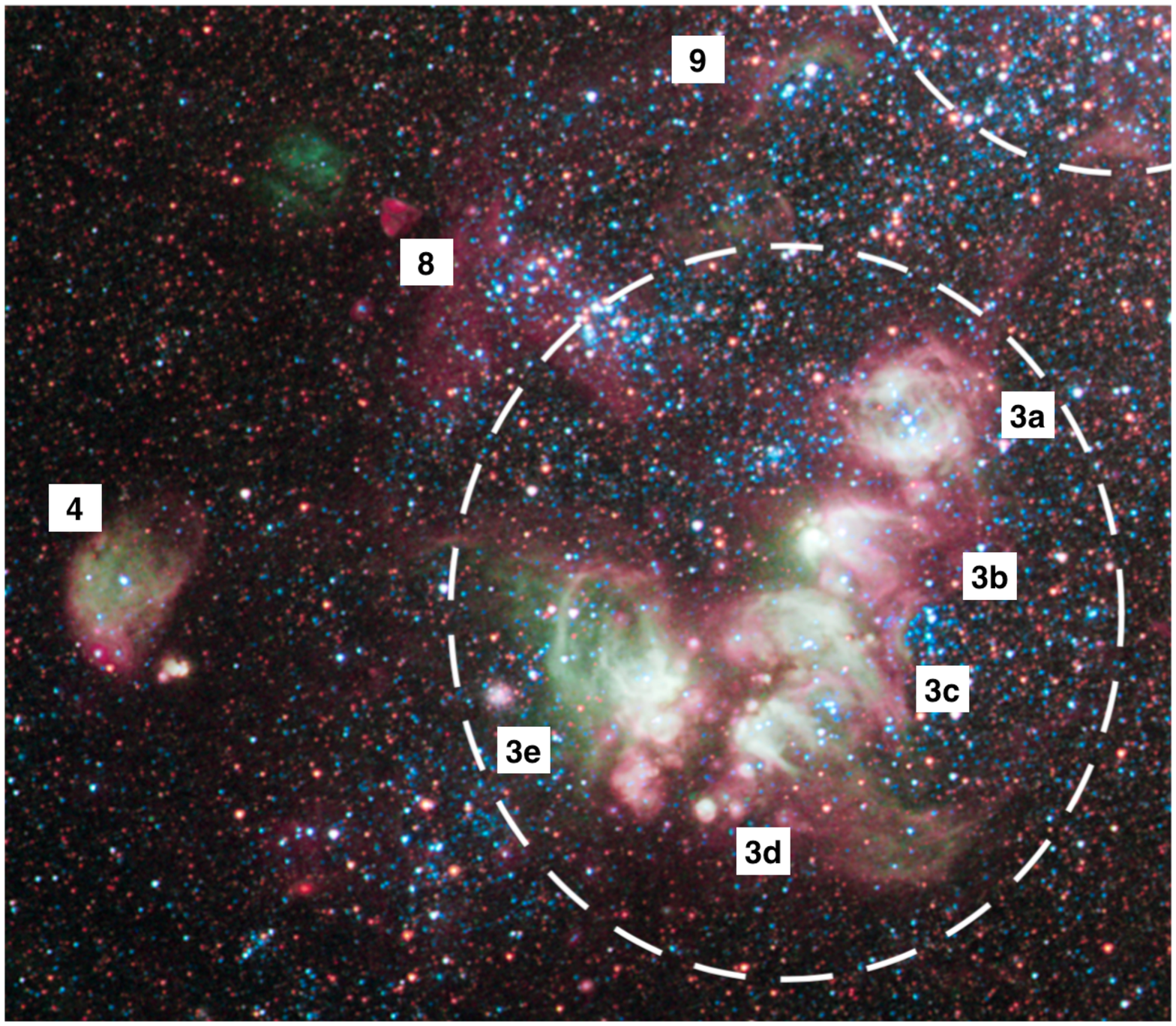}
\caption{{\it Left:} Map of the LyC escape fraction. The underlying gray-scale shows the extinction-corrected H$\alpha$ emission at the full HST/WFC3 resolution. Solid lines denote the individual SF regions, and are color-coded by their measured escape fractions. The local escape fraction varies from 0\% to 40\%. {\it Right:} A zoomed in view of around the SE SF complex. This includes individual SF regions 3a--3e, 4, 8, and 9.  
\label{mapEsc}}
\end{figure*}

\subsection{Local Escape Fraction Map for NGC 4214}
\label{sec:localEsc}
To properly measure the local $f_{\rm esc}$, we need to correctly associate stars with the local gas and dust that their ionizing photons interact with. We, therefore, conduct our analysis on individual star-forming regions. Since the boundary of each ionized HII region is a good tracer of a SF region's extent, individual SF regions should capture both the local production of ionizing photons and their interaction with the surrounding gas and dust. The locations and names of the 13 adopted SF regions can be found in Figure~\ref{mapEsc}, as well as in the extinction-corrected H$\alpha$ image in Figure~\ref{ancillary}.

Figure~\ref{mapEsc} presents the resulting local $f_{\rm esc}$ map for NGC 4214. For each region, we calculate $f_{\rm esc}$ from Eq.~\ref{eq_fesc}, using parameters measured within the SF region as listed in Table~\ref{tb_fesc}. The associated uncertainty is computed from the uncertainties in $\dot{N}_{\rm int}$ and $f_{\rm cov}$. Different boundary colors represent $f_{\rm esc}$ values for individual SF regions. The local escape fractions vary, with some regions having essentially no LyC leakage, and others having $f_{\rm esc}$ as large as 40\%. The escape fractions are high in the superbubbles, i.e., the SF regions 1 (40\%) and 2 (20\%), which together produce about 40\% of the total ionizing photons. Morphologically, these regions have clear shell structures that indicate the presence of large low-density holes, consistent with their high escape fractions. The SF region 9, located right below the Central SF complex, also shows a high escape fraction of 26\%. The SF regions 6 and 7 are other regions with high $f_{\rm esc}$ of 37\% and 26\%, respectively. They seem to have little gas and dust around the cluster of young stars while having significant FUV and H$\alpha$ emission. 

The SE complex (regions 3a--3e) is known to harbor the youngest stellar populations and highest H$\alpha$ surface brightness in NGC 4214 \citep[$\sim$2~Myr;][]{ubeda07b, hermelo13}, though the region is more obscured with a measured $f_{\rm cov}$ of 0.6. Both the young age and high $f_{\rm cov}$ of this complex indicate that it has not evolved long enough to decouple the stars from the birth clouds via significant stellar feedback (see also Section~\ref{sec:mass_temp}). Although the SE complex generally appears to be compact in H$\alpha$ emission, there are sub-regions (3a, 3d, and 3e) that are releasing non-negligible amounts of LyC photons ($\sim$6.16$\times$10$^{50}$~s$^{-1}$ together), likely through low-density holes with small opening angles. The escape fractions for 3b and 3c are 0\% and 10\%, respectively. These low $f_{\rm esc}$ regions show higher dust column densities along the line of sight and harbor a smaller number of detected ionizing sources, suggesting the possibility of presence of highly embedded stars. As mentioned earlier, our data are not complete for highly reddened stars with $A_{\rm V} >$ 2--3, possibly underpredicting the true intrinsic LyC photons in these SF regions. 

In addition to the uncertainties in the SED modeling and adopted $f_{\rm cov}$, there are two additional, but likely minor uncertainties: the effects of non-photoionized diffuse gas in the galactic halo and dust scattering. \citet{calzetti04} investigate the H$\alpha$ emission contribution from non-photoionized components to the total H$\alpha$ using emission lines for NGC 4214. The measured the fraction of non-photoionized gas, produced by shocks from stellar feedback, was $\sim$4\% of the total H$\alpha$ emission, well below our 20\% uncertainty on $\dot{N}_{\rm gas}$. The dust extinction we measure by modeling the observed SEDs includes both absorption and  scattering. A fraction of scattered light depends on many aspects such as dust column density, dust scattering properties, star/dust geometry, wavelength, etc. According to the detailed studies of dust scattering radiative transfer \citep[e.g.,][]{witt00}, the fraction of scattered light at the wavelengths of LyC would be insignificant for both the MW-type and SMC-type dust models with more realistic star/dust geometry (i.e., clumpy cloudy or clumpy dusty), particularly in a low optical depth environment. Therefore, we neglect the uncertainties due to these two effects.

\begin{deluxetable}{lcccc}
\tablecaption{Local Escape Fraction for Each SF Region \label{tb_fesc}} 
\tablewidth{0pt}

\tablehead{
\colhead{SF Region} & \colhead{$f_{\rm cov}$\tablenotemark{a}} & \colhead{f$_{\rm esc}$} & \colhead{$\dot{N}_{\rm int}$} &  \colhead{$\dot{N}_{\rm gas}$\tablenotemark{b}} \\
 & & (\%) &($10^{50}~s^{-1}$) &($10^{50}~s^{-1}$)}
\startdata
 1 (shell) & 0.3$^{+0.15}_{-0.1}$   &40$^{+10}_{-7}$   & 106.72$^{+6.24}_{-5.86}$\tablenotemark{c} &  45.69       \\
 2 (shell)  & 0.3$^{+0.15}_{-0.1}$   & 20$^{+7}_{-6}$   & 50.79$^{+4.14}_{-3.82}$  &  36.12    \\ 
 3a (small hole) & 0.6$^{+0.05}_{-0.3}$   & 19$^{+5}_{-18}$ & 9.48$^{+0.86}_{-2.41}$  & 5.09    \\
 3b (compact)  &  0.6$^{+0.05}_{-0.3}$ & 0\tablenotemark{d}  & 4.95  & 4.95      \\ 
 3c (compact)  &  0.6$^{+0.05}_{-0.3}$   & 10$^{+11}_{-10}$  & 5.18$^{+1.48}_{-0.78}$ & 3.84   \\
 3d (compact)  &  0.6$^{+0.05}_{-0.3}$   & 24$^{+12}_{-21}$   & 11.80$^{+2.76}_{-2.64}$  & 4.69   \\
 3e (compact)  &  0.6$^{+0.05}_{-0.3}$   &  13$^{+10}_{-11}$   & 11.75$^{+2.58}_{-1.58}$  & 8.05  \\
 4 (compact)   & 0.4$^{+0.2}_{-0.2}$     & 12$^{+12}_{-21}$  & 2.04$^{+0.36}_{-0.68}$ & 1.62  \\
 5 (compact)   & 0.4$^{+0.2}_{-0.2}$     &  14$^{+16}_{-16}$  & 2.43$^{+0.59}_{-0.62}$ & 1.88   \\
 6 (compact)   & 0.4$^{+0.2}_{-0.2}$    & 37$^{+15}_{-15}$  & 8.99$^{+0.99}_{-1.01}$  & 3.47   \\
 7 (compact)   & 0.4$^{+0.2}_{-0.2}$   & 26$^{+9}_{-9}$   & 4.51$^{+0.17}_{-0.12}$ & 2.55 \\
 8 (weak)   & 0.4$^{+0.2}_{-0.2}$    & 7$^{+10}_{-9}$  & 3.26$^{+0.55}_{-0.43}$  & 2.89   \\
 9 (weak)   & 0.3$^{+0.15}_{-0.1}$   & 26$^{+13}_{-13}$   & 2.85$^{+0.47}_{-0.48}$  & 1.81    \\
Diffuse\tablenotemark{e} & 0.475$^{+0.125}_{-0.125}$ & \ldots & 167.52$^{+6.82}_{-6.83}$ & 35.57 \\
\enddata
\tablecomments{The escape fraction of ionizing photons and adopted covering factor for each SF regions in NGC 4214.}
\tablenotetext{a}{The covering factor and its uncertainty from \citet{hermelo13}.}
\tablenotetext{b}{Associated uncertainties are $\sim$20\%.}
\tablenotetext{c}{This includes the contribution from the unresolved super star cluster I-As.}
\tablenotetext{d}{The BEAST predicts $\dot{N}_{\rm int} =$ 2.66$^{+0.81}_{-0.68}\times10^{50}$~s$^{-1}$, resulting in the escape fraction of --34\%. By assuming that all LyC photons contribute to photoionize the neutral hydrogen, we replace $\dot{N}_{\rm int}$ with $\dot{N}_{\rm gas}$.}
\tablenotemark{e}{Rest of area outside the 13 SF regions.}
\end{deluxetable}

\subsection{Global Escape Fraction of NGC 4214}
\label{sec:globalEsc}
Before the ionizing photons eventually leave a galaxy, they first have to escape from their birth clouds and then have to pass through the diffuse ISM without interaction. However, at least a fraction of ionizing photons will inevitably interact with the diffuse ISM in reality on the way out of the galaxy. 

Thus, the ingredients to obtain a global escape fraction for the entire galaxy are (1) the number of ionizing photons emerging from the SF regions ($\dot{N}_{\rm int}^{\rm SF}$), (2) the number of ionizing photons produced by stars outside the SF regions ($\dot{N}_{\rm int}^{\rm diffuse}$), (3) the number of ionizing photons consumed by the diffuse H$\alpha$ ($\dot{N}_{\rm gas}^{\rm diffuse}$), and (4) the covering factor of the diffuse dust ($f_{\rm cov}^{\rm diffuse}$). First, we combine the intrinsic ionizing photons escaping from the 13 SF regions and the intrinsic ionizing photons produced outside these SF regions. Second, we compute $\dot{N}_{\rm gas}$ powering the diffuse H$\alpha$, which is calculated by subtracting the combined $\dot{N}_{\rm gas}$ from the 13 SF regions from the total integrated $\dot{N}_{\rm gas}$. Finally, we use the diffuse dust $f_{\rm cov}^{\rm diffuse}$ of 0.35--0.60 with a mean of 0.475 \citep{hermelo13}. 

We consider the two diffuse ISM geometries in measuring the global $f_{\rm esc}$: (1) the diffuse ISM surrounded by a porous shell of dust, and (2) galactic outflows created in each SF region allow the LyC photons to escape directly to the IGM. The former case is the same ISM geometry we assume for individual SF region, and makes LyC photons first interaction with hydrogen in the diffuse ISM. In the first case, the global $f_{\rm esc}$ is computed as follows: 
\begin{equation} \label{eq_gloablfesc}
\begin{split}
f_{\rm esc}^{\rm global} & = \frac{
     (1 - f_{\rm cov}^{\rm diffuse})\times(\dot{N}_{\rm int}^{\rm SF} + \dot{N}_{\rm int}^{\rm diffuse} - \dot{N}_{\rm gas}^{\rm diffuse})}
  {\dot{N}_{\rm int}^{\rm stars}},
\end{split}
\end{equation}
where $\dot{N}_{\rm int}^{\rm stars}$ is the total production rate of LyC photons from all stars in NGC 4214. 

On the other hand, the latter case assumes that a fraction of LyC photons leaves the galaxy without interaction with the diffuse ISM and some of the residual LyC photons is consumed by photoionization in the diffuse ISM. In the latter case, the global $f_{\rm esc}$ is computed as follows: 
\begin{equation} \label{eq_gloablfesc}
\begin{split}
f_{\rm esc}^{\rm global} & = \frac{
     (1 - f_{\rm cov}^{\rm diffuse})\times(\dot{N}_{\rm int}^{\rm SF} + \dot{N}_{\rm int}^{\rm diffuse}) - \dot{N}_{\rm gas}^{\rm diffuse}}
  {\dot{N}_{\rm int}^{\rm stars}}.
\end{split}
\end{equation}

Our fiducial set of the local $f_{\rm cov}$ values are listed in Table~\ref{tb_fesc} (i.e., 0.3 for the SF regions 1, 2, and 9, 0.6 for the SF regions 3a--3e, 0.4 for the SF regions 4--8), and we adopted $f_{\rm cov} =$ 0.475 for the diffuse dust outside the SF regions. These choices give $f_{\rm esc} \simeq$ 27\% and 22\% for the diffuse ISM geometry case I and case II, respectively. We take the average of these two as our fiducial value of the global $f_{\rm esc} =$ 25\%. Based on the uncertainties in the local $f_{\rm cov}$ and $\dot{N}_{\rm int}$ listed in Table~\ref{tb_fesc}, 20\% uncertainty in $\dot{N}_{\rm gas}$, and the diffuse ISM geometry, we compute the lower and upper limits of the global escape fraction, which are 10\% and 41\%, respectively. Thus, the final global escape fraction is $f_{\rm esc}$ = \tol[25][16][15]\%.

Our fiducial $f_{\rm esc} =$ 25\% is the highest value measured for local analogs so far. With $M_{\rm FUV} =$ --15.9~mag, this result is consistent with a trend of increasing $f_{\rm esc}$ with decreasing UV luminosity seen in a high-resolution cosmological simulation \citep[e.g.,][]{anderson17}. Since our sample is nearly, but not 100\% complete for ionizing stars, our $\dot{N}_{\rm int}$ measurement might be slightly underestimated. This leads to overestimation in $f_{\rm esc}$ when $\dot{N}_{\rm gas}$ and $f_{\rm cov}$ remain unchanged. Our analysis shows that the global $f_{\rm esc}$ for NGC 4214 cannot be close to zero unless a majority of the ionizing stars were missed in our sample, which is not the case as we showed in Section~\ref{fuv_lyc}, or a significant amount of dust was undetected, which is hardly the case for the extremely well studied galaxy NGC 4214. 

\citet{lee09} derived the star formation rate (SFR) based on H$\alpha$ and UV, and found a ratio of SFR(H$\alpha$)/SFH(FUV) = 0.71 for NGC 4214. One possible cause for lowering the SFR based on H$\alpha$ is the leakage of ionizing photons. If we assume the LyC leakage is the true culprit, their result suggests that about 30\% of ionizing photons escapes from NGC 4214, which is consistent with our result. 

To our knowledge, $f_{\rm esc}$ for NGC 4214 has been reported in two studies up to date \citep{heckman01, grimes09}. They placed the upper limit on $f_{\rm esc}$ based on the residual intensity in the core of saturated CII $\lambda\,$1036 absorption line in the global spectra under the assumption that neutral carbon and hydrogen share the same $f_{\rm cov}$. Their results are basically consistent with zero escape fraction in our direction, but we note that this does not necessarily mean no escaping LyC photons in other directions. This is because detection of a residual flux near the center of absorption line also depends on viewing angle in the same manner to detection of leaking LyC. If neutral hydrogen indeed shares the same $f_{\rm cov}$ with neutral carbon, their findings indicate that there are no escaping LyC photons toward us because the opening angle of the low density holes does not align with our viewing angle.

\subsubsection{Escape from a Dust-free NGC 4214}
We also calculate the global escape fraction for the dust-free case. The dust-free (i.e., considering neutral hydrogen gas only) $f_{\rm esc}$ value provides an approximated value for NGC 4214-like galaxies at the epoch of reionization. A dwarf star-forming galaxy in the early Universe would have lower metallicity than NGC 4214, leading to lower dust content \citep[e.g.,][]{inoue16}. Due to the lower dust content in these galaxies, the dust effect on the ionizing photon escape fraction might be marginal. Thus, our assumption on the extreme dust-free case is not totally unrealistic for dwarf star-forming galaxies at the epoch of reionization. The global escape fraction for the dust-free case is measured to be $\sim$59\%, releasing $\sim$2.32$\times$10$^{52}$ ionizing photons per second. This is at least 10$\times$ higher than the inferred ionizing emissivity at $z >$ 6 \citep{bouwens15}. Recently, \citet{laporte17} reports the presence of dust in a star-forming galaxy at $z=$ 8.38 with a stellar mass of $\sim$2$\times$10$^9$\msun. A couple of additional galaxies at $z =$ 7--8 have been found to contain dust \citep{watson15, hashimoto19, tamura19}. Therefore, it is plausible for a high-$z$ counterpart of NGC 4214 with a small amount of dust to provide sufficient ionizing photons per Mpc$^3$ at the epoch of reionization.

\subsection{Effect of Stellar Rotation and Binary Stars}\label{sec:rotbinary}
The SED models used in this study do not include stellar rotation and binary evolution, which affect the ionizing photon production. Both rotation and binary evolution make the LyC regime of model SEDs harder, producing more hydrogen ionizing photons for a longer period of time \citep[e.g.,][]{topping15, stanway16, wilkins16, choi17}. Specifically, stellar models including the effect of interacting binary stars produce 60\% more ionizing flux at low metallicities ($<$ 0.3~\zsun) over an extended period compared to models without binaries. This is because, with mass and angular momentum transfer, the primary star has a harder ionizing spectrum and the secondary star becomes more massive allowing more massive stars to exist at later ages in an interacting binary system \citep{stanway16,eldridge17}. Theoretical studies have explored the effect of binary stars on the LyC escape fraction using cosmological zoom-in simulations and showed the increase in the escape fraction \citep[e.g.,][]{ma16,rosdahl18,ma20}. However, the extent to which the escape fraction increases remains a matter of debate. For example, \citet{ma16,rosdahl18} suggested the significant effect of binary stars (i.e., $f_{\rm esc}$ is boosted by at least 300\%), while \citet{ma20} reported a modest effect (i.e., $f_{\rm esc}$ is boosted by 25-35\%).    

With no changes in the observed SEDs from NUV to NIR, the dust extinction measurements based on the NUV to NIR would remain the same. Thus, for the same dust $f_{\rm cov}$ and $\dot{N}_{\rm gas}$, the inclusion of the stellar rotation and interacting binary effects will only increase the intrinsic LyC photon production rate, and thus eventually result in a higher escape fraction. In conclusion, our escape fraction measurement, which does not consider the stellar rotation and binary effects, sets a lower limit for NGC 4214. Taking the results by \citet{ma20} at face value, about a 25\% increase in $f_{\rm esc}$ is expected for a stellar mass of $\sim$10$^9$~\msun.

\section{Summary} \label{sec:summary}
We have developed a new technique for measuring the ionizing photon escape fraction with high spatial resolution. With the aid of multi-wavelength (NUV, optical, and NIR) broad-band imaging obtained by HST, we have modeled SEDs of resolved stars within NGC 4214 to infer their intrinsic stellar properties, as well as the intervening dust. We have used these to predict the non-ionizing FUV emission from the galaxy. We find excellent agreement with GALEX FUV measurements, giving us confidence in our estimates of LyC flux as well. We employ these fits to construct maps of ionizing photon production across the galaxy.

We have also measured the number of photons absorbed through photoionization using hydrogen emission lines in HST narrow-band imaging, and through dust by adopting the dust covering factor. We combine the emission and absorption measurements to map the escape fraction across the galaxy. This new technique can be used for any well-resolved galaxy without requiring spatially averaged, simplified assumptions about dust/ISM geometry along the line of sight for dust correction to properly measure the intrinsic LyC photon production rate, and without challenging observations below the Lyman limit. 

We find significant spatial variation in the local $f_{\rm esc}$ from 0\% to 40\%. We also find a broad range of the global $f_{\rm esc}$ with a fiducial value of $\sim$25\%, which is higher than the previous measurements from \citet{grimes09} by a factor of 125 and from \citet{heckman01} by a factor of 6.25, respectively, highlighting the importance of sufficient spatial resolution to resolve the relative star/dust geometry at small scales, and better considering 4$\pi$~sr ISM properties in the escape fraction measurement. Moreover, our results suggest that metal-poor UV-faint star-forming dwarf galaxies might have indeed provided sufficient number of ionizing photons to reionize the IGM at the epoch of reionization. 

While this current work provides the ionizing photon escape fraction map for NGC 4214 for the first time, we still need a larger sample of galaxies to obtain a detailed understanding of the mechanism for escape of ionizing photons and its relationship between the escape fraction and galactic environments (e.g., galaxy mass, star formation intensity, gas column density, metallicity). Detailed analyses of the local variation of the escape fraction are essential for understanding the degree to which physical properties of individual galaxies determine their contribution to cosmic reionization. Studies of lower metallicity galaxies are especially relevant in this regard.

\acknowledgements
We are grateful to the referee for providing constructive suggestions to improve the paper. We thank Knut Olsen for fruitful discussions. Support for this work was provided by NASA through grant number HST-AR-14288 from the Space Telescope Science Institute, which is operated by AURA, Inc., under NASA contract NAS 5-26555.

\software{scipy \citep{jones01}, numpy \citep{vanderwalt11}, matplotlib \citep{hunter07}, ipython \citep{PER-GRA:2007}, and astropy \citep{astropy18}}

\bibliographystyle{aasjournal}
\bibliography{references}
\clearpage

\end{document}